\documentclass[10pt,twocolumn]{article}

\usepackage[numbers]{natbib}
\usepackage[dvipsnames]{xcolor}
\usepackage{fullpage}
\usepackage{times}
\usepackage{colortbl}
\PassOptionsToPackage{hyphens}{url}\usepackage[colorlinks=false,hidelinks]{hyperref}

\usepackage{geometry}
\usepackage{pgfplots}
\pgfplotsset{compat=1.9}

\usepackage{subcaption}
\usepackage{graphicx}
\usepackage{layouts}
\usepackage{multirow}
\usepackage{xtab,booktabs}
\usepackage{longtable}
\usepackage{multicol}
\usepackage[inline]{enumitem} %
\usepackage{totpages}

\author{
David G. Balash\\
\small \tt david.balash@richmond.edu\\
\small \it University of Richmond
\and 
Mir Masood Ali\\
\small \tt mali92@uic.edu\\
\small \it University of Illinois Chicago
\and 
Monica Kodwani\\
\small \tt monicakodwani@gwu.edu\\
\small \it The George Washington University
\and 
Xiaoyuan Wu\\
\small \tt wxyowen@cmu.edu\\
\small \it Carnegie Mellon University
\and
Chris Kanich\\
\small \tt ckanich@uic.edu\\
\small \it University of Illinois Chicago
\and
Adam J. Aviv\\
\small \tt aaviv@gwu.edu\\
\small \it The George Washington University
}

\title{Longitudinal Analysis of Privacy Labels in the Apple App Store}
\date{\footnotesize Last Update: \today}

\newcommand{\changes}[1]{{\color{black}{#1}}} 

\newcommand{\pt}[0]{\emph{Privacy Type}}
\newcommand{\pu}[0]{\emph{Purpose}}
\newcommand{\pus}[0]{\emph{Purposes}}
\newcommand{\dc}[0]{\emph{Data Category}}

\newcommand{\pts}[0]{\emph{Privacy Types}}
\newcommand{\dcs}[0]{\emph{Data Categories}}
\newcommand{\dts}[0]{\emph{Data Types}}
\newcommand{\duty}[0]{\emph{Data Used to Track You}}
\newcommand{\dly}[0]{\emph{Data Linked to You}}
\newcommand{\dnly}[0]{\emph{Data Not Linked to You}}
\newcommand{\dnc}[0]{\emph{Data Not Collected}}

\newlist{questions}{enumerate}{3}
\setlist[questions]{align=left, labelwidth=2em, labelsep=.5em, listparindent=0pt, itemindent=0pt, leftmargin=!, before=\let\item\fixedItem}
\setlist[questions, 1]{labelindent=0pt, label=\textbf{Q\arabic*}, widest=99}
\setlist[questions, 2]{labelindent=-2.5em, label*=\textbf{\_\Alph*}, widest=26}
\setlist[questions, 3]{labelindent=-2.5em, label*=\textbf{\_\roman*}, widest=9}

\begin{document}
\newcommand{\appCostsStatisticsByPrivacyTypeTable}[0]{
\begin{table}[ht]
\centering
\caption{The number of \dcs~ collected per app by \pt~ statistics for free and paid apps.\\ ($M$ = median, $\mu$ = mean, $\sigma$ = standard deviation)}
\label{tab:app-costs-stats}

\begin{tabular}{l r r r | r r r}
\toprule
  \multicolumn{1}{l}{} &
  \multicolumn{3}{c}{\textbf{Free Apps}} &
  \multicolumn{3}{c}{\textbf{Paid Apps}} \\
  \multicolumn{1}{l}{} &
  \multicolumn{1}{c}{$M$} &
  \multicolumn{1}{c}{$\mu$} &
  \multicolumn{1}{c}{$\sigma$} &
  \multicolumn{1}{c}{$M$} &
  \multicolumn{1}{c}{$\mu$} &
  \multicolumn{1}{c}{$\sigma$}  \\
\hline
\rowcolor[HTML]{EFEFEF}
\textcolor[HTML]{d7191c}{\textbf{Used to Track}}       &   2  &  2.26 &  1.56  &  2 &  2.13  & 1.37\\
\textcolor[HTML]{fdae61}{\textbf{Data Linked}}        &	 4  &  6.16 &  6.22  &  3 &  4.85  & 4.80\\
\rowcolor[HTML]{EFEFEF}
\textcolor[HTML]{abd9e9}{\textbf{Data Not Linked}}      &   2  &  3.63 &  3.49  &  2 &  2.79  & 2.58\\
\textbf{Total}                                   &   5 	&  7.33 &  6.87  &  3 &  4.15  & 4.54\\
\bottomrule
\end{tabular}
\end{table}
}

\newcommand{\appCostsByPrivacyLabelTable}[0]{
\begin{table}[ht]
\centering
\caption{The app attributes by \pt.}
\label{tab:app-costs}

\begin{tabular}{l r r r r r}
\toprule
  \multicolumn{1}{l}{\textbf{}} &
  \multicolumn{1}{c}{\textbf{Num. Apps}} &
  \multicolumn{1}{c}{\textcolor[HTML]{d7191c}{\textbf{Data Used to}}} &
  \multicolumn{1}{c}{\textcolor[HTML]{fdae61}{\textbf{Data Linked}}} &
  \multicolumn{1}{c}{\textcolor[HTML]{abd9e9}{\textbf{Data Not}}} &
  \multicolumn{1}{c}{\textcolor[HTML]{2c7bb6}{\textbf{Data Not}}} \\
  \multicolumn{1}{l}{\textbf{}} &
  \multicolumn{1}{c}{\textbf{with Labels}} &
  \multicolumn{1}{c}{\textcolor[HTML]{d7191c}{\textbf{Track You}}} &
  \multicolumn{1}{c}{\textcolor[HTML]{fdae61}{\textbf{to You}}} &
  \multicolumn{1}{c}{\textcolor[HTML]{abd9e9}{\textbf{Linked to You}}} &
  \multicolumn{1}{c}{\textcolor[HTML]{2c7bb6}{\textbf{Collected}}} \\
\hline
\rowcolor[HTML]{EFEFEF}
\textbf{Free}                 & 785,320 &   110,316  & 304,056 & 315,211  & 337,536 \\
\rowcolor[HTML]{EFEFEF}
\textbf{} && (14.0\%) & (38.7\%) & (40.1\%) & (43.0\%)  \\
\textbf{Free In-App}          & 127,092 &	53,478  &  52,481 &  77,169  &	30,704 \\
\textbf{} && (42.1\%) & (41.3\%) & (60.7\%) & (24.2\%)  \\
\rowcolor[HTML]{EFEFEF}
\textbf{Paid}                & 36,980   &	   885  &	1,537 &   5,199  &	31,032 \\
\rowcolor[HTML]{EFEFEF}
\textbf{} && (2.4\%)  & (4.2\%)  & (14.1\%) & (83.9\%) \\
\textbf{Paid In-App}         &  3,530   &	   478 	&     551 &	  1,293  &	 1,981 \\
\textbf{} && (13.5\%) & (15.6\%) & (36.6\%) & (56.1\%) \\
\hline
\rowcolor[HTML]{EFEFEF}
\textbf{100kB}       & \multicolumn{1}{r}{6,406}   & 80    & 588    & 647    & 5,411   \\
\rowcolor[HTML]{EFEFEF} 
\textbf{}            &                            & (1.2\%)   & (9.2\%)    & (10.1\%)   & (84.5\%)   \\
\textbf{1MB}         & \multicolumn{1}{r}{62,667}  & 4,714  & 10,456  & 13,967  & 43,567  \\
\textbf{}            &                            & (7.5\%)   & (16.7\%)   & (22.3\%)   & (69.5\%)   \\
\rowcolor[HTML]{EFEFEF}
\textbf{10MB}        & \multicolumn{1}{r}{679,668} & 98,810 & 259,203 & 281,771 & 287,705 \\
\rowcolor[HTML]{EFEFEF}
\textbf{}            &                            & (14.5\%)  & (38.1\%)   & (41.5\%)   & (42.3\%)   \\
\textbf{100MB}       & \multicolumn{1}{r}{199,527} & 59,607 & 86,720  & 99,977  & 63,060  \\
\textbf{}            &                            & (29.9\%)  & (43.5\%)   & (50.1\%)   & (31.6\%)   \\
\rowcolor[HTML]{EFEFEF}
\textbf{1GB}         & \multicolumn{1}{r}{4,276}   & 1,737  & 1,530   & 2,274   & 1,411   \\
\rowcolor[HTML]{EFEFEF}
\textbf{}            &                            & (40.6\%)  & (35.8\%)   & (53.2\%)   & (33.0\%) \\

\bottomrule
\end{tabular}
\end{table}
}

\newcommand{\privacyTypeShiftsWithVersionUpdates}[0]{
\begin{table*}[ht]
\centering
\normalsize
\caption{Privacy Type Shifts in correlation with version updates.}
\label{tab:privacy-type-shifts-with-version-updates}
\begin{tabular}{@{}lcrr@{}}
\toprule
\textbf{Initial Posture} & \textbf{New Posture} & \textbf{Without Version Update} & \textbf{With Version Update} \\ \midrule
\cellcolor[HTML]{2C7BB6}    & \cellcolor[HTML]{ABD9E9}Data Not Linked to You & 1,769 & 2,629 \\
\cellcolor[HTML]{2C7BB6}    & \cellcolor[HTML]{FDAE61}Data Linked to You     & 1,433 & 1,960 \\
\multirow{-3}{*}{\cellcolor[HTML]{2C7BB6}Data Not Collected}    & \cellcolor[HTML]{D7191C}Data Used to Track You & 1,066 & 1,918 \\
\hline
\cellcolor[HTML]{ABD9E9}  & \cellcolor[HTML]{2C7BB6}Data Not Collected     & 281  & 685  \\
\cellcolor[HTML]{ABD9E9} & \cellcolor[HTML]{FDAE61}Data Linked to You     & 416  & 769  \\
\multirow{-3}{*}{\cellcolor[HTML]{ABD9E9}Data Not Linked to You} & \cellcolor[HTML]{D7191C}Data Used to Track You & 219  & 364  \\
\hline
\cellcolor[HTML]{FDAE61}    & \cellcolor[HTML]{2C7BB6}Data Not Collected     & 246  & 572  \\
\cellcolor[HTML]{FDAE61}    & \cellcolor[HTML]{ABD9E9}Data Not Linked to You & 586  & 681  \\
\multirow{-3}{*}{\cellcolor[HTML]{FDAE61}Data Linked to You}     & \cellcolor[HTML]{D7191C}Data Used to Track You & 286  & 246  \\
\hline
\cellcolor[HTML]{D7191C}    & \cellcolor[HTML]{2C7BB6}Data Not Collected     & 286  & 815  \\
\cellcolor[HTML]{D7191C}    & \cellcolor[HTML]{ABD9E9}Data Not Linked to You & 1,322 & 2,585 \\
\multirow{-3}{*}{\cellcolor[HTML]{D7191C}Data Used to Track You}    & \cellcolor[HTML]{FDAE61}Data Linked to You     & 1,353 & 2,701 \\
\bottomrule
\end{tabular}%
\end{table*}
}

\newcommand{\dataCategoryShiftsLabeledTable}[0]{
\begin{table}[t]
\centering
\caption{Changes observed in the addition and removal of data categories from privacy labels. Twice as many apps that changed the composition of their labels added a data category as those that removed a data category.}
\label{tab:data-category-shifts-labeled}
\normalsize
\begin{tabular}{l r r}
\toprule
\multicolumn{1}{l}{\textbf{Data Category}} &
  \multicolumn{1}{c}{\textcolor[HTML]{2c7bb6}{\textbf{Added}}} &
  \multicolumn{1}{c}{\textcolor[HTML]{d7191c}{\textbf{Removed}}} \\
\hline
\rowcolor[HTML]{EFEFEF}
Identifiers          & 8,576 & 2,585 \\
Usage Data          & 7,317 & 1,993 \\
\rowcolor[HTML]{EFEFEF}
Diagnostics          & 6,208 & 1,811 \\
Contact Info        & 5,439 & 1,765 \\
\rowcolor[HTML]{EFEFEF}
Location             & 5,489 & 1900 \\
User Content        & 3,688 & 1,444 \\
\rowcolor[HTML]{EFEFEF}
Purchases            & 1,705 & 784  \\
Other                & 1,505 & 760  \\
\rowcolor[HTML]{EFEFEF}
Financial Info      & 1,781 & 654  \\
Search History      & 897  & 613  \\
\rowcolor[HTML]{EFEFEF}
Contacts             & 572  & 326  \\
Browsing History    & 472  & 427  \\
\rowcolor[HTML]{EFEFEF}
Health and Fitness & 395  & 139  \\
Sensitive Info      & 300  & 147 \\
\bottomrule
\end{tabular}%
\end{table}
}

\newcommand{\privacyTypeShiftsWithinDataCategory}[0]{
\onecolumn
\begin{longtable}[c]{@{}lccrrc@{}}
\caption{Observed shifts in privacy types associated with each data category.}
\label{tab:privacy-type-shifts-data-categories}\\
\toprule
\textbf{\begin{tabular}[c]{@{}c@{}}Data\\Category\end{tabular}} &
  \textbf{\begin{tabular}[c]{@{}c@{}}From \\ Privacy \\ Type\end{tabular}} &
  \textbf{\begin{tabular}[c]{@{}c@{}}To \\ Privacy \\ Type\end{tabular}} &
  \textbf{\begin{tabular}[c]{@{}c@{}}Shift \\ Count\end{tabular}} &
  \textbf{\begin{tabular}[c]{@{}c@{}}Total \\ Shifts\end{tabular}} &
  \textbf{\begin{tabular}[c]{@{}c@{}}Prominent\\App\\Genre\end{tabular}} \\* \midrule
\endfirsthead
\multicolumn{6}{c}%
{{\bfseries Table \thetable\ continued from previous page}} \\
\toprule
\textbf{\begin{tabular}[c]{@{}c@{}}Data\\Category\end{tabular}} &
  \textbf{\begin{tabular}[c]{@{}c@{}}From \\ Privacy \\ Type\end{tabular}} &
  \textbf{\begin{tabular}[c]{@{}c@{}}To \\ Privacy \\ Type\end{tabular}} &
  \textbf{\begin{tabular}[c]{@{}c@{}}Shift \\ Count\end{tabular}} &
  \textbf{\begin{tabular}[c]{@{}c@{}}Total \\ Shifts\end{tabular}} &
  \textbf{\begin{tabular}[c]{@{}c@{}}Prominent\\App\\Genre\end{tabular}} \\* \midrule
\endhead
 &
 \cellcolor[HTML]{ABD9E9}Data Not Linked to You &
  \cellcolor[HTML]{D7191C}Data Used to Track You &
  \cellcolor[HTML]{EFEFEF}2,489 &
   &
   \\* \cmidrule(lr){2-4}
 &
  \cellcolor[HTML]{FDAE61}Data Linked to You &
  \cellcolor[HTML]{D7191C}Data Used to Track You &
  946 &
   &
   \\* \cmidrule(lr){2-4}
 &
  \cellcolor[HTML]{ABD9E9}Data Not Linked to You &
  \cellcolor[HTML]{FDAE61}Data Linked to You &
  \cellcolor[HTML]{EFEFEF}1,205 &
   &
   \\* \cmidrule(lr){2-4}
 &
  \cellcolor[HTML]{D7191C}Data Used to Track You &
  \cellcolor[HTML]{FDAE61}Data Linked to You &
  183 &
   &
   \\* \cmidrule(lr){2-4}
 &
  \cellcolor[HTML]{FDAE61}Data Linked to You &
  \cellcolor[HTML]{ABD9E9}Data Not Linked to You &
  1,042 &
   &
   \\* \cmidrule(lr){2-4}
\multirow{-6}{*}{Usage Data} &
  \cellcolor[HTML]{D7191C}Data Used to Track You &
  \cellcolor[HTML]{ABD9E9}Data Not Linked to You &
  548 &
  \multirow{-6}{*}{6,413} &
  \multirow{-6}{*}{Games} \\* \midrule
 &
  \cellcolor[HTML]{ABD9E9}Data Not Linked to You &
  \cellcolor[HTML]{D7191C}Data Used to Track You &
  \cellcolor[HTML]{EFEFEF}1480 &
   &
   \\* \cmidrule(lr){2-4}
 &
  \cellcolor[HTML]{FDAE61}Data Linked to You &
  \cellcolor[HTML]{D7191C}Data Used to Track You &
  \cellcolor[HTML]{EFEFEF}1,742 &
   &
   \\* \cmidrule(lr){2-4}
 &
  \cellcolor[HTML]{ABD9E9}Data Not Linked to You &
  \cellcolor[HTML]{FDAE61}Data Linked to You &
  916 &
   &
   \\* \cmidrule(lr){2-4}
 &
 \cellcolor[HTML]{D7191C}Data Used to Track You &
  \cellcolor[HTML]{FDAE61}Data Linked to You &
  232 &
   &
   \\* \cmidrule(lr){2-4}
 &
  \cellcolor[HTML]{FDAE61}Data Linked to You &
  \cellcolor[HTML]{ABD9E9}Data Not Linked to You &
  1,016 &
   &
   \\* \cmidrule(lr){2-4}
\multirow{-6}{*}{Identifiers} &
  \cellcolor[HTML]{D7191C}Data Used to Track You &
  \cellcolor[HTML]{ABD9E9}Data Not Linked to You &
  518
  &
  \multirow{-6}{*}{5,904} &
  \multirow{-6}{*}{Games} \\* \midrule

 &
  \cellcolor[HTML]{ABD9E9}Data Not Linked to You &
  \cellcolor[HTML]{D7191C}Data Used to Track You &
  \cellcolor[HTML]{EFEFEF}590 &
   &
   \\* \cmidrule(lr){2-4}
 &
  \cellcolor[HTML]{FDAE61}Data Linked to You &
  \cellcolor[HTML]{D7191C}Data Used to Track You &
  154 &
   &
   \\* \cmidrule(lr){2-4}
 &
  \cellcolor[HTML]{ABD9E9}Data Not Linked to You &
  \cellcolor[HTML]{FDAE61}Data Linked to You &
  \cellcolor[HTML]{EFEFEF}666 &
   &
   \\* \cmidrule(lr){2-4}
 &
  \cellcolor[HTML]{D7191C}Data Used to Track You &
  \cellcolor[HTML]{FDAE61}Data Linked to You &
  55 &
   &
   \\* \cmidrule(lr){2-4}
 &
  \cellcolor[HTML]{FDAE61}Data Linked to You &
  \cellcolor[HTML]{ABD9E9}Data Not Linked to You &
  468 &
   &
   \\* \cmidrule(lr){2-4}
\multirow{-6}{*}{Diagnostics} &
  \cellcolor[HTML]{D7191C}Data Used to Track You &
  \cellcolor[HTML]{ABD9E9}Data Not Linked to You &
  244 &
  \multirow{-6}{*}{2,177} &
  \multirow{-6}{*}{Games} \\* \midrule
 &
  \cellcolor[HTML]{ABD9E9}Data Not Linked to You &
  \cellcolor[HTML]{D7191C}Data Used to Track You &
  \cellcolor[HTML]{EFEFEF}405 &
   &
   \\* \cmidrule(lr){2-4}
 &
  \cellcolor[HTML]{FDAE61}Data Linked to You &
  \cellcolor[HTML]{D7191C}Data Used to Track You &
  241 &
   &
   \\* \cmidrule(lr){2-4}
 &
  \cellcolor[HTML]{ABD9E9}Data Not Linked to You &
  \cellcolor[HTML]{FDAE61}Data Linked to You &
  398 &
   &
   \\* \cmidrule(lr){2-4}
 &
  \cellcolor[HTML]{D7191C}Data Used to Track You &
  \cellcolor[HTML]{FDAE61}Data Linked to You &
  47 &
   &
   \\* \cmidrule(lr){2-4}
 &
  \cellcolor[HTML]{FDAE61}Data Linked to You &
  \cellcolor[HTML]{ABD9E9}Data Not Linked to You &
  \cellcolor[HTML]{EFEFEF}578 &
   &
   \\* \cmidrule(lr){2-4}
\multirow{-6}{*}{Location} &
  \cellcolor[HTML]{D7191C}Data Used to Track You &
  \cellcolor[HTML]{ABD9E9}Data Not Linked to You &
  344 &
  \multirow{-6}{*}{2,013} &
  \multirow{-6}{*}{Games} \\* \midrule
 &
  \cellcolor[HTML]{ABD9E9}Data Not Linked to You &
  \cellcolor[HTML]{D7191C}Data Used to Track You &
  170 &
   &
   \\* \cmidrule(lr){2-4}
 &
  \cellcolor[HTML]{FDAE61}Data Linked to You &
  \cellcolor[HTML]{D7191C}Data Used to Track You &
  \cellcolor[HTML]{EFEFEF}581 &
   &
   \\* \cmidrule(lr){2-4}
 &
  \cellcolor[HTML]{ABD9E9}Data Not Linked to You &
  \cellcolor[HTML]{FDAE61}Data Linked to You &
  415 &
   &
   \\* \cmidrule(lr){2-4}
 &
  \cellcolor[HTML]{D7191C}Data Used to Track You &
  \cellcolor[HTML]{FDAE61}Data Linked to You &
  44 &
   &
   \\* \cmidrule(lr){2-4}
 &
  \cellcolor[HTML]{FDAE61}Data Linked to You &
  \cellcolor[HTML]{ABD9E9}Data Not Linked to You &
  \cellcolor[HTML]{EFEFEF}507 &
   &
   \\* \cmidrule(lr){2-4}
\multirow{-6}{*}{Contact Info} &
  \cellcolor[HTML]{D7191C}Data Used to Track You &
  \cellcolor[HTML]{ABD9E9}Data Not Linked to You &
  255 &
  \multirow{-6}{*}{1,972} &
  \multirow{-6}{*}{Shopping} \\* \midrule

 \pagebreak
 &
  \cellcolor[HTML]{ABD9E9}Data Not Linked to You &
  \cellcolor[HTML]{D7191C}Data Used to Track You &
  62 &
   &
   \\* \cmidrule(lr){2-4}
 &
  \cellcolor[HTML]{FDAE61}Data Linked to You &
  \cellcolor[HTML]{D7191C}Data Used to Track You &
  112 &
   &
   \\* \cmidrule(lr){2-4}
 &
  \cellcolor[HTML]{ABD9E9}Data Not Linked to You &
  \cellcolor[HTML]{FDAE61}Data Linked to You &
  \cellcolor[HTML]{EFEFEF}238 &
   &
   \\* \cmidrule(lr){2-4}
 &
  \cellcolor[HTML]{D7191C}Data Used to Track You &
  \cellcolor[HTML]{FDAE61}Data Linked to You &
  6 &
   &
   \\* \cmidrule(lr){2-4}
 &
  \cellcolor[HTML]{FDAE61}Data Linked to You &
  \cellcolor[HTML]{ABD9E9}Data Not Linked to You &
  \cellcolor[HTML]{EFEFEF}311 &
   &
   \\* \cmidrule(lr){2-4}
\multirow{-6}{*}{User Content} &
  \cellcolor[HTML]{D7191C}Data Used to Track You &
  \cellcolor[HTML]{ABD9E9}Data Not Linked to You &
  96 &
  \multirow{-6}{*}{825} &
  \multirow{-6}{*}{Business} \\* \midrule
 &
  \cellcolor[HTML]{ABD9E9}Data Not Linked to You &
  \cellcolor[HTML]{D7191C}Data Used to Track You &
  62 &
   &
   \\* \cmidrule(lr){2-4}
 &
  \cellcolor[HTML]{FDAE61}Data Linked to You &
  \cellcolor[HTML]{D7191C}Data Used to Track You &
  \cellcolor[HTML]{EFEFEF}148 &
   &
   \\* \cmidrule(lr){2-4}
 &
  \cellcolor[HTML]{ABD9E9}Data Not Linked to You &
  \cellcolor[HTML]{FDAE61}Data Linked to You &
  82 &
   &
   \\* \cmidrule(lr){2-4}
 &
  \cellcolor[HTML]{D7191C}Data Used to Track You &
  \cellcolor[HTML]{FDAE61}Data Linked to You &
  13 &
   &
   \\* \cmidrule(lr){2-4}
 &
  \cellcolor[HTML]{FDAE61}Data Linked to You &
  \cellcolor[HTML]{ABD9E9}Data Not Linked to You &
  \cellcolor[HTML]{EFEFEF}103 &
   &
   \\* \cmidrule(lr){2-4}
\multirow{-6}{*}{Purchases} &
  \cellcolor[HTML]{D7191C}Data Used to Track You &
  \cellcolor[HTML]{ABD9E9}Data Not Linked to You &
  49 &
  \multirow{-6}{*}{457} &
  \multirow{-6}{*}{Games} \\* \midrule
 &
  \cellcolor[HTML]{ABD9E9}Data Not Linked to You &
  \cellcolor[HTML]{D7191C}Data Used to Track You &
  41 &
   &
   \\* \cmidrule(lr){2-4}
 &
  \cellcolor[HTML]{FDAE61}Data Linked to You &
  \cellcolor[HTML]{D7191C}Data Used to Track You &
  17 &
   &
   \\* \cmidrule(lr){2-4}
 &
  \cellcolor[HTML]{ABD9E9}Data Not Linked to You &
  \cellcolor[HTML]{FDAE61}Data Linked to You &
  \cellcolor[HTML]{EFEFEF}190 &
   &
   \\* \cmidrule(lr){2-4}
 &
  \cellcolor[HTML]{D7191C}Data Used to Track You &
  \cellcolor[HTML]{FDAE61}Data Linked to You &
  4 &
   &
   \\* \cmidrule(lr){2-4}
 &
  \cellcolor[HTML]{FDAE61}Data Linked to You &
  \cellcolor[HTML]{ABD9E9}Data Not Linked to You &
  \cellcolor[HTML]{EFEFEF}107 &
   &
   \\* \cmidrule(lr){2-4}
\multirow{-6}{*}{Search History} &
  \cellcolor[HTML]{D7191C}Data Used to Track You &
  \cellcolor[HTML]{ABD9E9}Data Not Linked to You &
  32 &
  \multirow{-6}{*}{391} &
  \multirow{-6}{*}{Shopping} \\* \midrule
 
 &
  \cellcolor[HTML]{ABD9E9}Data Not Linked to You &
  \cellcolor[HTML]{D7191C}Data Used to Track You &
  \cellcolor[HTML]{EFEFEF}53 &
   &
   \\* \cmidrule(lr){2-4}
 &
  \cellcolor[HTML]{FDAE61}Data Linked to You &
  \cellcolor[HTML]{D7191C}Data Used to Track You &
  32 &
   &
   \\* \cmidrule(lr){2-4}
 &
  \cellcolor[HTML]{ABD9E9}Data Not Linked to You &
  \cellcolor[HTML]{FDAE61}Data Linked to You &
  37 &
   &
   \\* \cmidrule(lr){2-4}
 &
  \cellcolor[HTML]{D7191C}Data Used to Track You &
  \cellcolor[HTML]{FDAE61}Data Linked to You &
  4 &
   &
   \\* \cmidrule(lr){2-4}
 &
  \cellcolor[HTML]{FDAE61}Data Linked to You &
  \cellcolor[HTML]{ABD9E9}Data Not Linked to You &
  \cellcolor[HTML]{EFEFEF}76 &
   &
   \\* \cmidrule(lr){2-4}
\multirow{-6}{*}{Other} &
  \cellcolor[HTML]{D7191C}Data Used to Track You &
  \cellcolor[HTML]{ABD9E9}Data Not Linked to You &
  51 &
  \multirow{-6}{*}{253} &
  \multirow{-8}{*}{\begin{tabular}[c]{@{}c@{}}Games \\ Business \\ Health\\\& Fitness\end{tabular}} \\* \midrule
 &
  \cellcolor[HTML]{ABD9E9}Data Not Linked to You &
  \cellcolor[HTML]{D7191C}Data Used to Track You &
  6 &
   &
   \\* \cmidrule(lr){2-4}
 &
  \cellcolor[HTML]{FDAE61}Data Linked to You &
  \cellcolor[HTML]{D7191C}Data Used to Track You &
  18 &
   &
   \\* \cmidrule(lr){2-4}
 &
  \cellcolor[HTML]{ABD9E9}Data Not Linked to You &
  \cellcolor[HTML]{FDAE61}Data Linked to You &
  \cellcolor[HTML]{EFEFEF}38 &
   &
   \\* \cmidrule(lr){2-4}
 &
  \cellcolor[HTML]{D7191C}Data Used to Track You &
  \cellcolor[HTML]{FDAE61}Data Linked to You &
  2 &
   &
   \\* \cmidrule(lr){2-4}
 &
  \cellcolor[HTML]{FDAE61}Data Linked to You &
  \cellcolor[HTML]{ABD9E9}Data Not Linked to You &
  \cellcolor[HTML]{EFEFEF}65 &
   &
   \\* \cmidrule(lr){2-4}
\multirow{-6}{*}{Financial Info} &
  \cellcolor[HTML]{D7191C}Data Used to Track You &
  \cellcolor[HTML]{ABD9E9}Data Not Linked to You &
  25 &
  \multirow{-6}{*}{154} &
  \multirow{-6}{*}{Finance} \\* \midrule
 \pagebreak
  &
  \cellcolor[HTML]{ABD9E9}Data Not Linked to You &
  \cellcolor[HTML]{D7191C}Data Used to Track You &
  \cellcolor[HTML]{EFEFEF}37 &
   &
   \\* \cmidrule(lr){2-4}
 &
  \cellcolor[HTML]{FDAE61}Data Linked to You &
  \cellcolor[HTML]{D7191C}Data Used to Track You &
  12 &
   &
   \\* \cmidrule(lr){2-4}
 &
  \cellcolor[HTML]{ABD9E9}Data Not Linked to You &
  \cellcolor[HTML]{FDAE61}Data Linked to You &
  12 &
   &
   \\* \cmidrule(lr){2-4}
 &
  \cellcolor[HTML]{D7191C}Data Used to Track You &
  \cellcolor[HTML]{FDAE61}Data Linked to You &
  7 &
   &
   \\* \cmidrule(lr){2-4}
 &
  \cellcolor[HTML]{FDAE61}Data Linked to You &
  \cellcolor[HTML]{ABD9E9}Data Not Linked to You &
  \cellcolor[HTML]{EFEFEF}63 &
   &
   \\* \cmidrule(lr){2-4}
\multirow{-7}{*}{\begin{tabular}[c]{@{}c@{}}Browsing\\History\end{tabular}} &
  \cellcolor[HTML]{D7191C}Data Used to Track You &
  \cellcolor[HTML]{ABD9E9}Data Not Linked to You &
  21 &
  \multirow{-6}{*}{152} &
  \multirow{-6}{*}{Shopping} \\* \midrule
 &
  \cellcolor[HTML]{ABD9E9}Data Not Linked to You &
  \cellcolor[HTML]{D7191C}Data Used to Track You &
  5 &
   &
   \\* \cmidrule(lr){2-4}
 &
  \cellcolor[HTML]{FDAE61}Data Linked to You &
  \cellcolor[HTML]{D7191C}Data Used to Track You &
  2 &
   &
   \\* \cmidrule(lr){2-4}
 &
  \cellcolor[HTML]{ABD9E9}Data Not Linked to You &
  \cellcolor[HTML]{FDAE61}Data Linked to You &
  \cellcolor[HTML]{EFEFEF}21 &
   &
   \\* \cmidrule(lr){2-4}
 &
  \cellcolor[HTML]{D7191C}Data Used to Track You &
  \cellcolor[HTML]{FDAE61}Data Linked to You &
  1 &
   &
   \\* \cmidrule(lr){2-4}
 &
  \cellcolor[HTML]{FDAE61}Data Linked to You &
  \cellcolor[HTML]{ABD9E9}Data Not Linked to You &
  \cellcolor[HTML]{EFEFEF}38 &
   &
   \\* \cmidrule(lr){2-4}
\multirow{-6}{*}{Contacts} &
  \cellcolor[HTML]{D7191C}Data Used to Track You &
  \cellcolor[HTML]{ABD9E9}Data Not Linked to You &
  15 &
  \multirow{-6}{*}{82} &
  \multirow{-6}{*}{Finance} \\* \midrule
 &
  \cellcolor[HTML]{ABD9E9}Data Not Linked to You &
  \cellcolor[HTML]{D7191C}Data Used to Track You &
  3 &
   &
   \\* \cmidrule(lr){2-4}
 &
  \cellcolor[HTML]{FDAE61}Data Linked to You &
  \cellcolor[HTML]{D7191C}Data Used to Track You &
  2 &
   &
   \\* \cmidrule(lr){2-4}
 &
  \cellcolor[HTML]{ABD9E9}Data Not Linked to You &
  \cellcolor[HTML]{FDAE61}Data Linked to You &
  \cellcolor[HTML]{EFEFEF}25 &
   &
   \\* \cmidrule(lr){2-4}
 &
  \cellcolor[HTML]{D7191C}Data Used to Track You &
  \cellcolor[HTML]{FDAE61}Data Linked to You &
  2 &
   &
   \\* \cmidrule(lr){2-4}
 &
  \cellcolor[HTML]{D7191C}Data Used to Track You &
  \cellcolor[HTML]{ABD9E9}Data Not Linked to You &
  2 &
   &
   \\* \cmidrule(lr){2-4}
\multirow{-8}{*}{\begin{tabular}[c]{@{}c@{}}Health\\\&\\Fitness\end{tabular}} &
  \cellcolor[HTML]{FDAE61}Data Linked to You &
  \cellcolor[HTML]{ABD9E9}Data Not Linked to You &
  \cellcolor[HTML]{EFEFEF}18 &
  \multirow{-6}{*}{52} &
  \multirow{-9}{*}{\begin{tabular}[c]{@{}c@{}}Medical\\and\\Health\\\&\\Fitness\end{tabular}} \\* \midrule

 &
  \cellcolor[HTML]{ABD9E9}Data Not Linked to You &
  \cellcolor[HTML]{D7191C}Data Used to Track You &
  4 &
   &
   \\* \cmidrule(lr){2-4}
 &
  \cellcolor[HTML]{FDAE61}Data Linked to You &
  \cellcolor[HTML]{D7191C}Data Used to Track You &
  4 &
   &
   \\* \cmidrule(lr){2-4}
 &
  \cellcolor[HTML]{ABD9E9}Data Not Linked to You &
  \cellcolor[HTML]{FDAE61}Data Linked to You &
  \cellcolor[HTML]{EFEFEF}13 &
   &
   \\* \cmidrule(lr){2-4}
 &
  \cellcolor[HTML]{D7191C}Data Used to Track You &
  \cellcolor[HTML]{FDAE61}Data Linked to You &
  2 &
   &
   \\* \cmidrule(lr){2-4}
 &
  \cellcolor[HTML]{FDAE61}Data Linked to You &
  \cellcolor[HTML]{ABD9E9}Data Not Linked to You &
  \cellcolor[HTML]{EFEFEF}13 &
   &
   \\* \cmidrule(lr){2-4}
\multirow{-6}{*}{\begin{tabular}[c]{@{}c@{}}Sensitive\\Info\end{tabular}} &
  \cellcolor[HTML]{D7191C}Data Used to Track You &
  \cellcolor[HTML]{ABD9E9}Data Not Linked to You &
  3 &
  \multirow{-6}{*}{39} &
  \multirow{-6}{*}{\begin{tabular}[c]{@{}c@{}}Health\\\&\\Fitness\end{tabular}} \\* \bottomrule
\end{longtable}
\twocolumn
}

\newcommand{\purposeShiftsWithinDataCategory}[0]{
\onecolumn
\begin{longtable}[c]{lllrr}
\caption{Observed shifts in purposes associated with each data category.}
\label{tab:purpose-shifts-data-categories}\\
\hline
\multicolumn{1}{c}{\textbf{Data Category}} &
  \multicolumn{1}{c}{\textbf{From Purpose}} &
  \multicolumn{1}{c}{\textbf{To Purpose}} &
  \multicolumn{1}{c}{\textbf{Count}} &
  \textbf{Total No. of Shifts} \\ \hline
\endfirsthead
\multicolumn{5}{c}%
{{\bfseries Table \thetable\ continued from previous page}} \\
\hline
\multicolumn{1}{c}{\textbf{Data Category}} &
  \multicolumn{1}{c}{\textbf{From Purpose}} &
  \multicolumn{1}{c}{\textbf{To Purpose}} &
  \multicolumn{1}{c}{\textbf{Count}} &
  \textbf{Total No. of Shifts} \\ \hline
\endhead
\cline{2-4}
\endfoot
\endlastfoot
 &
  \cellcolor[HTML]{EA4335}App Functionality &
  \cellcolor[HTML]{EA4335}Analytics &
  \cellcolor[HTML]{EA4335}777 &
   \\
 &
  Analytics &
  App Functionality &
  318 &
   \\
 &
  \cellcolor[HTML]{F4CCCC}App Functionality &
  \cellcolor[HTML]{F4CCCC}Third Party Adveritising &
  \cellcolor[HTML]{F4CCCC}642 &
   \\
 &
  Third Party Adveritising &
  App Functionality &
  147 &
   \\
 &
  App Functionality &
  Product Personalization &
  569 &
   \\
 &
  Product Personalization &
  App Functionality &
  71 &
   \\
 &
  App Functionality &
  Developers Advertising &
  559 &
   \\
 &
  Developers Advertising &
  App Functionality &
  87 &
   \\
 &
  App Functionality &
  Other Purposes &
  176 &
   \\
 &
  Other Purposes &
  App Functionality &
  58 &
   \\
 &
  Analytics &
  Third Party Adveritising &
  494 &
   \\
 &
  Third Party Adveritising &
  Analytics &
  329 &
   \\
 &
  Analytics &
  Product Personalization &
  450 &
   \\
 &
  Product Personalization &
  Analytics &
  147 &
   \\
 &
  Analytics &
  Developers Advertising &
  431 &
   \\
 &
  Developers Advertising &
  Analytics &
  73 &
   \\
 &
  Analytics &
  Other Purposes &
  109 &
   \\
 &
  Other Purposes &
  Analytics &
  47 &
   \\
 &
  Third Party Adveritising &
  Product Personalization &
  118 &
   \\
 &
  Product Personalization &
  Third Party Adveritising &
  198 &
   \\
 &
  Third Party Adveritising &
  Developers Advertising &
  109 &
   \\
 &
  Developers Advertising &
  Third Party Adveritising &
  123 &
   \\
 &
  Third Party Adveritising &
  Other Purposes &
  39 &
   \\
 &
  Other Purposes &
  Third Party Adveritising &
  33 &
   \\
 &
  Product Personalization &
  Developers Advertising &
  188 &
   \\
 &
  Developers Advertising &
  Product Personalization &
  147 &
   \\
 &
  Product Personalization &
  Other Purposes &
  66 &
   \\
 &
  Other Purposes &
  Product Personalization &
  23 &
   \\
 &
  Developers Advertising &
  Other Purposes &
  33 &
   \\
\multirow{-30}{*}{Identifiers} &
  Other Purposes &
  Developers Advertising &
  39 &
  \multirow{-30}{*}{6,
  600} \\
\hline
 &
  App Functionality &
  Analytics &
  371 &
   \\
 &
  Analytics &
  App Functionality &
  539 &
   \\
 &
  App Functionality &
  Third Party Adveritising &
  443 &
   \\
 &
  Third Party Adveritising &
  App Functionality &
  143 &
   \\
 &
  App Functionality &
  Product Personalization &
  312 &
   \\
 &
  Product Personalization &
  App Functionality &
  91 &
   \\
 &
  App Functionality &
  Developers Advertising &
  306 &
   \\
 &
  Developers Advertising &
  App Functionality &
  94 &
   \\
 &
  App Functionality &
  Other Purposes &
  110 &
   \\
 &
  Other Purposes &
  App Functionality &
  56 &
   \\
 &
  \cellcolor[HTML]{EA4335}Analytics &
  \cellcolor[HTML]{EA4335}Third Party Adveritising &
  \cellcolor[HTML]{EA4335}1,040 &
   \\
 &
  Third Party Adveritising &
  Analytics &
  204 &
   \\
\multirow{-7}{*}{Usage Data} &
  Analytics &
  Product Personalization &
  439 &
  \multirow{-7}{*}{6,206}
   \\
 &
  Product Personalization &
  Analytics &
  62 &
   \\
 &
  \cellcolor[HTML]{F4CCCC}Analytics &
  \cellcolor[HTML]{F4CCCC}Developers Advertising &
  \cellcolor[HTML]{F4CCCC}588 &
   \\
 &
  Developers Advertising &
  Analytics &
  37 &
   \\
 &
  Analytics &
  Other Purposes &
  151 &
   \\
 &
  Other Purposes &
  Analytics &
  111 &
   \\
 &
  Third Party Adveritising &
  Product Personalization &
  107 &
   \\
 &
  Product Personalization &
  Third Party Adveritising &
  213 &
   \\
 &
  Third Party Adveritising &
  Developers Advertising &
  124 &
   \\
 &
  Developers Advertising &
  Third Party Adveritising &
  145 &
   \\
 &
  Third Party Adveritising &
  Other Purposes &
  64 &
   \\
 &
  Other Purposes &
  Third Party Adveritising &
  41 &
   \\
 &
  Product Personalization &
  Developers Advertising &
  131 &
   \\
 &
  Developers Advertising &
  Product Personalization &
  114 &
   \\
 &
  Product Personalization &
  Other Purposes &
  61 &
   \\
 &
  Other Purposes &
  Product Personalization &
  25 &
   \\
 &
  Developers Advertising &
  Other Purposes &
  48 &
   \\
\multirow{-10}{*}{Usage Data} &
  Other Purposes &
  Developers Advertising &
  36 &
\\
\hline
 &
  \cellcolor[HTML]{EA4335}App Functionality &
  \cellcolor[HTML]{EA4335}Analytics &
  \cellcolor[HTML]{EA4335}454 &
   \\
 &
  \cellcolor[HTML]{F4CCCC}Analytics &
  \cellcolor[HTML]{F4CCCC}App Functionality &
  \cellcolor[HTML]{F4CCCC}413 &
   \\
 &
  App Functionality &
  Third Party Adveritising &
  155 &
   \\
 &
  Third Party Adveritising &
  App Functionality &
  67 &
   \\
 &
  App Functionality &
  Product Personalization &
  144 &
   \\
 &
  Product Personalization &
  App Functionality &
  13 &
   \\
 &
  App Functionality &
  Developers Advertising &
  49 &
   \\
 &
  Developers Advertising &
  App Functionality &
  15 &
   \\
 &
  App Functionality &
  Other Purposes &
  68 &
   \\
 &
  Other Purposes &
  App Functionality &
  123 &
   \\
 &
  Analytics &
  Third Party Adveritising &
  170 &
   \\
 &
  Third Party Adveritising &
  Analytics &
  63 &
   \\
 &
  Analytics &
  Product Personalization &
  88 &
   \\
 &
  Product Personalization &
  Analytics &
  22 &
   \\
 &
  Analytics &
  Developers Advertising &
  65 &
   \\
 &
  Developers Advertising &
  Analytics &
  10 &
   \\
 &
  Analytics &
  Other Purposes &
  90 &
   \\
 &
  Other Purposes &
  Analytics &
  58 &
   \\
 &
  Third Party Adveritising &
  Product Personalization &
  8 &
   \\
 &
  Product Personalization &
  Third Party Adveritising &
  19 &
   \\
 &
  Third Party Adveritising &
  Developers Advertising &
  8 &
   \\
 &
  Developers Advertising &
  Third Party Adveritising &
  6 &
   \\
 &
  Third Party Adveritising &
  Other Purposes &
  11 &
   \\
 &
  Other Purposes &
  Third Party Adveritising &
  13 &
   \\
 &
  Product Personalization &
  Developers Advertising &
  11 &
   \\
 &
  Developers Advertising &
  Product Personalization &
  9 &
   \\
 &
  Product Personalization &
  Other Purposes &
  3 &
   \\
 &
  Other Purposes &
  Product Personalization &
  13 &
   \\
 &
  Developers Advertising &
  Other Purposes &
  4 &
   \\
\multirow{-30}{*}{Diagnostics} &
  Other Purposes &
  Developers Advertising &
  9 &
  \multirow{-30}{*}{2,181} \\
\hline
 &
  \cellcolor[HTML]{F4CCCC}App Functionality &
  \cellcolor[HTML]{F4CCCC}Analytics &
  \cellcolor[HTML]{F4CCCC}307 &
   \\
 &
  Analytics &
  App Functionality &
  80 &
   \\
 &
  App Functionality &
  Third Party Adveritising &
  118 &
   \\
Contact Info &
  Third Party Adveritising &
  App Functionality &
  14 &
   2,409 \\
 &
  \cellcolor[HTML]{EA4335}App Functionality &
  \cellcolor[HTML]{EA4335}Product Personalization &
  \cellcolor[HTML]{EA4335}339 &
   \\
 &
  Product Personalization &
  App Functionality &
  52 &
   \\
 &
  App Functionality &
  Developers Advertising &
  298 &
   \\
 &
  Developers Advertising &
  App Functionality &
  52 &
   \\
 &
  App Functionality &
  Other Purposes &
  139 &
   \\
 &
  Other Purposes &
  App Functionality &
  85 &
   \\
 &
  Analytics &
  Third Party Adveritising &
  59 &
   \\
 &
  Third Party Adveritising &
  Analytics &
  8 &
   \\
 &
  Analytics &
  Product Personalization &
  84 &
   \\
 &
  Product Personalization &
  Analytics &
  97 &
   \\
 &
  Analytics &
  Developers Advertising &
  99 &
   \\
 &
  Developers Advertising &
  Analytics &
  53 &
   \\
 &
  Analytics &
  Other Purposes &
  44 &
   \\
 &
  Other Purposes &
  Analytics &
  44 &
   \\
 &
  Third Party Adveritising &
  Product Personalization &
  13 &
   \\
 &
  Product Personalization &
  Third Party Adveritising &
  56 &
   \\
 &
  Third Party Adveritising &
  Developers Advertising &
  16 &
   \\
 &
  Developers Advertising &
  Third Party Adveritising &
  48 &
   \\
 &
  Third Party Adveritising &
  Other Purposes &
  11 &
   \\
 &
  Other Purposes &
  Third Party Adveritising &
  16 &
   \\
 &
  Product Personalization &
  Developers Advertising &
  85 &
   \\
 &
  Developers Advertising &
  Product Personalization &
  45 &
   \\
 &
  Product Personalization &
  Other Purposes &
  46 &
   \\
 &
  Other Purposes &
  Product Personalization &
  53 &
   \\
 &
  Developers Advertising &
  Other Purposes &
  19 &
   \\
\multirow{-30}{*}{Contact Info} &
  Other Purposes &
  Developers Advertising &
  29 &
  \multirow{-30}{*}{2,409} \\
\hline
 &
  \cellcolor[HTML]{F4CCCC}App Functionality &
  \cellcolor[HTML]{F4CCCC}Analytics &
  \cellcolor[HTML]{F4CCCC}178 &
   \\
 &
  \cellcolor[HTML]{EA4335}Analytics &
  \cellcolor[HTML]{EA4335}App Functionality &
  \cellcolor[HTML]{EA4335}220 &
   \\
 &
  App Functionality &
  Third Party Adveritising &
  102 &
   \\
 &
  Third Party Adveritising &
  App Functionality &
  93 &
   \\
 &
  App Functionality &
  Product Personalization &
  142 &
   \\
 &
  Product Personalization &
  App Functionality &
  86 &
   \\
 &
  App Functionality &
  Developers Advertising &
  95 &
   \\
 &
  Developers Advertising &
  App Functionality &
  82 &
   \\
 &
  App Functionality &
  Other Purposes &
  74 &
   \\
 &
  Other Purposes &
  App Functionality &
  98 &
   \\
 &
  Analytics &
  Third Party Adveritising &
  76 &
   \\
 &
  Third Party Adveritising &
  Analytics &
  54 &
   \\
 &
  Analytics &
  Product Personalization &
  74 &
   \\
 &
  Product Personalization &
  Analytics &
  61 &
   \\
 &
  Analytics &
  Developers Advertising &
  78 &
   \\
 &
  Developers Advertising &
  Analytics &
  19 &
   \\
 Location &
  Analytics &
  Other Purposes &
  70 &
  1,944 \\
 &
  Other Purposes &
  Analytics &
  19 &
   \\
 &
  Third Party Adveritising &
  Product Personalization &
  27 &
   \\
 &
  Product Personalization &
  Third Party Adveritising &
  45 &
   \\
 &
  Third Party Adveritising &
  Developers Advertising &
  37 &
   \\
 &
  Developers Advertising &
  Third Party Adveritising &
  22 &
   \\
 &
  Third Party Adveritising &
  Other Purposes &
  17 &
   \\
 &
  Other Purposes &
  Third Party Adveritising &
  12 &
   \\
 &
  Product Personalization &
  Developers Advertising &
  53 &
   \\
 &
  Developers Advertising &
  Product Personalization &
  20 &
   \\
 &
  Product Personalization &
  Other Purposes &
  40 &
   \\
 &
  Other Purposes &
  Product Personalization &
  19 &
   \\
 &
  Developers Advertising &
  Other Purposes &
  23 &
   \\
\multirow{-2}{*}{Location} &
  Other Purposes &
  Developers Advertising &
  8 &
  \multirow{-2}{*}{1,944} \\
\hline
 &
  \cellcolor[HTML]{EA4335}App Functionality &
  \cellcolor[HTML]{EA4335}Analytics &
  \cellcolor[HTML]{EA4335}135 &
   \\
 &
  Analytics &
  App Functionality &
  48 &
   \\
 &
  App Functionality &
  Third Party Adveritising &
  37 &
   \\
 &
  Third Party Adveritising &
  App Functionality &
  9 &
   \\
 &
  \cellcolor[HTML]{F4CCCC}App Functionality &
  \cellcolor[HTML]{F4CCCC}Product Personalization &
  \cellcolor[HTML]{F4CCCC}108 &
   \\
 &
  Product Personalization &
  App Functionality &
  48 &
   \\
 &
  App Functionality &
  Developers Advertising &
  33 &
   \\
 &
  Developers Advertising &
  App Functionality &
  11 &
   \\
 &
  App Functionality &
  Other Purposes &
  69 &
   \\
 &
  Other Purposes &
  App Functionality &
  89 &
   \\
 &
  Analytics &
  Third Party Adveritising &
  11 &
   \\
 &
  Third Party Adveritising &
  Analytics &
  0 &
   \\
 &
  Analytics &
  Product Personalization &
  30 &
   \\
 &
  Product Personalization &
  Analytics &
  25 &
   \\
 &
  Analytics &
  Developers Advertising &
  15 &
   \\
 &
  Developers Advertising &
  Analytics &
  4 &
   \\
 &
  Analytics &
  Other Purposes &
  17 &
   \\
 &
  Other Purposes &
  Analytics &
  23 &
   \\
 &
  Third Party Adveritising &
  Product Personalization &
  1 &
   \\
 &
  Product Personalization &
  Third Party Adveritising &
  16 &
   \\
 &
  Third Party Adveritising &
  Developers Advertising &
  0 &
   \\
 &
  Developers Advertising &
  Third Party Adveritising &
  2 &
   \\
 &
  Third Party Adveritising &
  Other Purposes &
  2 &
   \\
 &
  Other Purposes &
  Third Party Adveritising &
  10 &
   \\
 &
  Product Personalization &
  Developers Advertising &
  13 &
   \\
 &
  Developers Advertising &
  Product Personalization &
  12 &
   \\
 &
  Product Personalization &
  Other Purposes &
  18 &
   \\
 &
  Other Purposes &
  Product Personalization &
  17 &
   \\
 &
  Developers Advertising &
  Other Purposes &
  6 &
   \\
\multirow{-30}{*}{User Content} &
  Other Purposes &
  Developers Advertising &
  16 &
  \multirow{-30}{*}{825} \\
\hline
 &
  \cellcolor[HTML]{EA4335}App Functionality &
  \cellcolor[HTML]{EA4335}Analytics &
  \cellcolor[HTML]{EA4335}64 &
   \\
 &
  Analytics &
  App Functionality &
  34 &
   \\
 &
  App Functionality &
  Third Party Adveritising &
  52 &
   \\
 &
  Third Party Adveritising &
  App Functionality &
  7 &
   \\
 &
  App Functionality &
  Product Personalization &
  49 &
   \\
 &
  Product Personalization &
  App Functionality &
  17 &
   \\
 &
  App Functionality &
  Developers Advertising &
  52 &
   \\
 &
  Developers Advertising &
  App Functionality &
  13 &
   \\
 &
  App Functionality &
  Other Purposes &
  12 &
   \\
 Purchases &
  Other Purposes &
  App Functionality &
  9 &
  604 \\
 &
  \cellcolor[HTML]{F4CCCC}Analytics &
  \cellcolor[HTML]{F4CCCC}Third Party Adveritising &
  \cellcolor[HTML]{F4CCCC}60 &
   \\
 &
  Third Party Adveritising &
  Analytics &
  8 &
   \\
 &
  Analytics &
  Product Personalization &
  31 &
   \\
 &
  Product Personalization &
  Analytics &
  16 &
   \\
 &
  Analytics &
  Developers Advertising &
  43 &
   \\
 &
  Developers Advertising &
  Analytics &
  8 &
   \\
 &
  Analytics &
  Other Purposes &
  8 &
   \\
 &
  Other Purposes &
  Analytics &
  7 &
   \\
 &
  Third Party Adveritising &
  Product Personalization &
  6 &
   \\
 &
  Product Personalization &
  Third Party Adveritising &
  37 &
   \\
 &
  Third Party Adveritising &
  Developers Advertising &
  4 &
   \\
 &
  Developers Advertising &
  Third Party Adveritising &
  20 &
   \\
 &
  Third Party Adveritising &
  Other Purposes &
  3 &
   \\
 &
  Other Purposes &
  Third Party Adveritising &
  6 &
   \\
 &
  Product Personalization &
  Developers Advertising &
  14 &
   \\
 &
  Developers Advertising &
  Product Personalization &
  4 &
   \\
 &
  Product Personalization &
  Other Purposes &
  9 &
   \\
 &
  Other Purposes &
  Product Personalization &
  6 &
   \\
 &
  Developers Advertising &
  Other Purposes &
  0 &
   \\
\multirow{-13}{*}{Purchases} &
  Other Purposes &
  Developers Advertising &
  5 &
  \multirow{-13}{*}{604} \\
\hline
 &
  \cellcolor[HTML]{EA4335}App Functionality &
  \cellcolor[HTML]{EA4335}Analytics &
  \cellcolor[HTML]{EA4335}21 &
   \\
 &
  Analytics &
  App Functionality &
  8 &
   \\
 &
  App Functionality &
  Third Party Adveritising &
  4 &
   \\
 &
  Third Party Adveritising &
  App Functionality &
  2 &
   \\
 &
  App Functionality &
  Product Personalization &
  16 &
   \\
 &
  Product Personalization &
  App Functionality &
  6 &
   \\
 &
  App Functionality &
  Developers Advertising &
  13 &
   \\
 &
  Developers Advertising &
  App Functionality &
  3 &
   \\
 &
  App Functionality &
  Other Purposes &
  16 &
   \\
 &
  \cellcolor[HTML]{F4CCCC}Other Purposes &
  \cellcolor[HTML]{F4CCCC}App Functionality &
  \cellcolor[HTML]{F4CCCC}17 &
   \\
 &
  Analytics &
  Third Party Adveritising &
  1 &
   \\
 &
  Third Party Adveritising &
  Analytics &
  2 &
   \\
 &
  Analytics &
  Product Personalization &
  3 &
   \\
 &
  Product Personalization &
  Analytics &
  3 &
   \\
 &
  Analytics &
  Developers Advertising &
  4 &
   \\
 &
  Developers Advertising &
  Analytics &
  1 &
   \\
 &
  Analytics &
  Other Purposes &
  4 &
   \\
 &
  Other Purposes &
  Analytics &
  5 &
   \\
 &
  Third Party Adveritising &
  Product Personalization &
  1 &
   \\
 &
  Product Personalization &
  Third Party Adveritising &
  0 &
   \\
 &
  Third Party Adveritising &
  Developers Advertising &
  0 &
   \\
 &
  Developers Advertising &
  Third Party Adveritising &
  0 &
   \\
 &
  Third Party Adveritising &
  Other Purposes &
  1 &
   \\
 &
  Other Purposes &
  Third Party Adveritising &
  2 &
   \\
 &
  Product Personalization &
  Developers Advertising &
  4 &
   \\
 &
  Developers Advertising &
  Product Personalization &
  1 &
   \\
 &
  Product Personalization &
  Other Purposes &
  5 &
   \\
 &
  Other Purposes &
  Product Personalization &
  8 &
   \\
 &
  Developers Advertising &
  Other Purposes &
  1 &
   \\
\multirow{-30}{*}{Financial Info} &
  Other Purposes &
  Developers Advertising &
  3 &
  \multirow{-30}{*}{155} \\
\hline
 &
  App Functionality &
  Analytics &
  16 &
   \\
 &
  \cellcolor[HTML]{EA4335}Analytics &
  \cellcolor[HTML]{EA4335}App Functionality &
  \cellcolor[HTML]{EA4335}25 &
   \\
 &
  App Functionality &
  Third Party Adveritising &
  14 &
   \\
 &
  Third Party Adveritising &
  App Functionality &
  9 &
   \\
 Other &
  App Functionality &
  Product Personalization &
  15 &
  284
   \\
 &
  Product Personalization &
  App Functionality &
  4 &
   \\
 &
  \cellcolor[HTML]{F4CCCC}App Functionality &
  \cellcolor[HTML]{F4CCCC}Developers Advertising &
  \cellcolor[HTML]{F4CCCC}20 &
   \\
 &
  Developers Advertising &
  App Functionality &
  0 &
   \\
 &
  App Functionality &
  Other Purposes &
  16 &
   \\
 &
  Other Purposes &
  App Functionality &
  15 &
   \\
 &
  Analytics &
  Third Party Adveritising &
  19 &
   \\
 &
  Third Party Adveritising &
  Analytics &
  3 &
   \\
 &
  Analytics &
  Product Personalization &
  10 &
   \\
 &
  Product Personalization &
  Analytics &
  3 &
   \\
 &
  Analytics &
  Developers Advertising &
  22 &
   \\
 &
  Developers Advertising &
  Analytics &
  0 &
   \\
 &
  Analytics &
  Other Purposes &
  12 &
   \\
 &
  Other Purposes &
  Analytics &
  7 &
   \\
 &
  Third Party Adveritising &
  Product Personalization &
  8 &
   \\
 &
  Product Personalization &
  Third Party Adveritising &
  7 &
   \\
 &
  Third Party Adveritising &
  Developers Advertising &
  8 &
   \\
 &
  Developers Advertising &
  Third Party Adveritising &
  3 &
   \\
 &
  Third Party Adveritising &
  Other Purposes &
  4 &
   \\
 &
  Other Purposes &
  Third Party Adveritising &
  13 &
   \\
 &
  Product Personalization &
  Developers Advertising &
  8 &
   \\
 &
  Developers Advertising &
  Product Personalization &
  1 &
   \\
 &
  Product Personalization &
  Other Purposes &
  7 &
   \\
 &
  Other Purposes &
  Product Personalization &
  7 &
   \\
 &
  Developers Advertising &
  Other Purposes &
  3 &
   \\
\multirow{-25}{*}{Other} &
  Other Purposes &
  Developers Advertising &
  5 &
  \multirow{-25}{*}{284} \\
\hline
 &
  App Functionality &
  Analytics &
  28 &
   \\
 &
  \cellcolor[HTML]{EA4335}Analytics &
  \cellcolor[HTML]{EA4335}App Functionality &
  \cellcolor[HTML]{EA4335}44 &
   \\
 &
  App Functionality &
  Third Party Adveritising &
  14 &
   \\
 &
  Third Party Adveritising &
  App Functionality &
  1 &
   \\
 &
  App Functionality &
  Product Personalization &
  23 &
   \\
 &
  Product Personalization &
  App Functionality &
  14 &
   \\
 &
  App Functionality &
  Developers Advertising &
  11 &
   \\
 &
  Developers Advertising &
  App Functionality &
  5 &
   \\
 &
  App Functionality &
  Other Purposes &
  4 &
   \\
 &
  Other Purposes &
  App Functionality &
  3 &
   \\
 &
  Analytics &
  Third Party Adveritising &
  18 &
   \\
 &
  Third Party Adveritising &
  Analytics &
  0 &
   \\
 &
  \cellcolor[HTML]{F4CCCC}Analytics &
  \cellcolor[HTML]{F4CCCC}Product Personalization &
  \cellcolor[HTML]{F4CCCC}36 &
   \\
 &
  Product Personalization &
  Analytics &
  15 &
   \\
 Search History &
  Analytics &
  Developers Advertising &
  18 &
  291 \\
 &
  Developers Advertising &
  Analytics &
  0 &
   \\
 &
  Analytics &
  Other Purposes &
  4 &
   \\
 &
  Other Purposes &
  Analytics &
  4 &
   \\
 &
  Third Party Adveritising &
  Product Personalization &
  0 &
   \\
 &
  Product Personalization &
  Third Party Adveritising &
  13 &
   \\
 &
  Third Party Adveritising &
  Developers Advertising &
  1 &
   \\
 &
  Developers Advertising &
  Third Party Adveritising &
  11 &
   \\
 &
  Third Party Adveritising &
  Other Purposes &
  0 &
   \\
 &
  Other Purposes &
  Third Party Adveritising &
  3 &
   \\
 &
  Product Personalization &
  Developers Advertising &
  8 &
   \\
 &
  Developers Advertising &
  Product Personalization &
  6 &
   \\
 &
  Product Personalization &
  Other Purposes &
  2 &
   \\
 &
  Other Purposes &
  Product Personalization &
  1 &
   \\
 &
  Developers Advertising &
  Other Purposes &
  1 &
   \\
\multirow{-2}{*}{Search History} &
  Other Purposes &
  Developers Advertising &
  3 &
  \multirow{-2}{*}{291} \\
\hline
 &
  App Functionality &
  Analytics &
  5 &
   \\
 &
  Analytics &
  App Functionality &
  5 &
   \\
 &
  App Functionality &
  Third Party Adveritising &
  1 &
   \\
 &
  Third Party Adveritising &
  App Functionality &
  1 &
   \\
 &
  \cellcolor[HTML]{EA4335}App Functionality &
  \cellcolor[HTML]{EA4335}Product Personalization &
  \cellcolor[HTML]{EA4335}90 &
   \\
 &
  Product Personalization &
  App Functionality &
  7 &
   \\
 &
  App Functionality &
  Developers Advertising &
  5 &
   \\
 &
  Developers Advertising &
  App Functionality &
  0 &
   \\
 &
  App Functionality &
  Other Purposes &
  5 &
   \\
 &
  Other Purposes &
  App Functionality &
  6 &
   \\
 &
  Analytics &
  Third Party Adveritising &
  2 &
   \\
 &
  Third Party Adveritising &
  Analytics &
  0 &
   \\
 &
  Analytics &
  Product Personalization &
  1 &
   \\
 &
  Product Personalization &
  Analytics &
  2 &
   \\
 &
  Analytics &
  Developers Advertising &
  2 &
   \\
 &
  Developers Advertising &
  Analytics &
  1 &
   \\
 &
  Analytics &
  Other Purposes &
  1 &
   \\
 &
  Other Purposes &
  Analytics &
  0 &
   \\
 &
  Third Party Adveritising &
  Product Personalization &
  0 &
   \\
 &
  Product Personalization &
  Third Party Adveritising &
  1 &
   \\
 &
  Third Party Adveritising &
  Developers Advertising &
  0 &
   \\
 &
  Developers Advertising &
  Third Party Adveritising &
  1 &
   \\
 &
  Third Party Adveritising &
  Other Purposes &
  0 &
   \\
 &
  Other Purposes &
  Third Party Adveritising &
  1 &
   \\
 &
  Product Personalization &
  Developers Advertising &
  1 &
   \\
 &
  Developers Advertising &
  Product Personalization &
  0 &
   \\
 &
  Product Personalization &
  Other Purposes &
  1 &
   \\
 &
  Other Purposes &
  Product Personalization &
  2 &
   \\
 &
  Developers Advertising &
  Other Purposes &
  0 &
   \\
\multirow{-30}{*}{Contacts} &
  Other Purposes &
  Developers Advertising &
  0 &
  \multirow{-30}{*}{141} \\
\hline
 &
  App Functionality &
  Analytics &
  14 &
   \\
 &
  Analytics &
  App Functionality &
  9 &
   \\
 &
  App Functionality &
  Third Party Adveritising &
  3 &
   \\
 &
  Third Party Adveritising &
  App Functionality &
  0 &
   \\
 &
  \cellcolor[HTML]{F4CCCC}App Functionality &
  \cellcolor[HTML]{F4CCCC}Product Personalization &
  \cellcolor[HTML]{F4CCCC}39 &
   \\
 &
  Product Personalization &
  App Functionality &
  9 &
   \\
 &
  App Functionality &
  Developers Advertising &
  4 &
   \\
 &
  Developers Advertising &
  App Functionality &
  0 &
   \\
 &
  App Functionality &
  Other Purposes &
  1 &
   \\
 Health \& Fitness &
  Other Purposes &
  App Functionality &
  4 &
  141
   \\
 &
  Analytics &
  Third Party Adveritising &
  0 &
   \\
 &
  Third Party Adveritising &
  Analytics &
  0 &
   \\
 &
  \cellcolor[HTML]{EA4335}Analytics &
  \cellcolor[HTML]{EA4335}Product Personalization &
  \cellcolor[HTML]{EA4335}43 &
   \\
 &
  Product Personalization &
  Analytics &
  4 &
   \\
 &
  Analytics &
  Developers Advertising &
  1 &
   \\
 &
  Developers Advertising &
  Analytics &
  0 &
   \\
 &
  Analytics &
  Other Purposes &
  2 &
   \\
 &
  Other Purposes &
  Analytics &
  2 &
   \\
 &
  Third Party Adveritising &
  Product Personalization &
  0 &
   \\
 &
  Product Personalization &
  Third Party Adveritising &
  0 &
   \\
 &
  Third Party Adveritising &
  Developers Advertising &
  0 &
   \\
 &
  Developers Advertising &
  Third Party Adveritising &
  0 &
   \\
 &
  Third Party Adveritising &
  Other Purposes &
  0 &
   \\
 &
  Other Purposes &
  Third Party Adveritising &
  1 &
   \\
 &
  Product Personalization &
  Developers Advertising &
  1 &
   \\
 &
  Developers Advertising &
  Product Personalization &
  0 &
   \\
 &
  Product Personalization &
  Other Purposes &
  1 &
   \\
 &
  Other Purposes &
  Product Personalization &
  2 &
   \\
 &
  Developers Advertising &
  Other Purposes &
  0 &
   \\
\multirow{-13}{*}{Health \& Fitness} &
  Other Purposes &
  Developers Advertising &
  1 &
  \multirow{-13}{*}{141} \\
\hline
 &
  \cellcolor[HTML]{EA4335}App Functionality &
  \cellcolor[HTML]{EA4335}Analytics &
  \cellcolor[HTML]{EA4335}22 &
   \\
 &
  Analytics &
  App Functionality &
  8 &
   \\
 &
  App Functionality &
  Third Party Adveritising &
  4 &
   \\
 &
  Third Party Adveritising &
  App Functionality &
  2 &
   \\
 &
  App Functionality &
  Product Personalization &
  3 &
   \\
 &
  Product Personalization &
  App Functionality &
  1 &
   \\
 &
  \cellcolor[HTML]{F4CCCC}App Functionality &
  \cellcolor[HTML]{F4CCCC}Developers Advertising &
  \cellcolor[HTML]{F4CCCC}21 &
   \\
 &
  Developers Advertising &
  App Functionality &
  4 &
   \\
 &
  App Functionality &
  Other Purposes &
  2 &
   \\
 &
  Other Purposes &
  App Functionality &
  0 &
   \\
 &
  Analytics &
  Third Party Adveritising &
  4 &
   \\
 &
  Third Party Adveritising &
  Analytics &
  1 &
   \\
 &
  Analytics &
  Product Personalization &
  4 &
   \\
 &
  Product Personalization &
  Analytics &
  1 &
   \\
 &
  Analytics &
  Developers Advertising &
  5 &
   \\
 &
  Developers Advertising &
  Analytics &
  0 &
   \\
 &
  Analytics &
  Other Purposes &
  4 &
   \\
 &
  Other Purposes &
  Analytics &
  1 &
   \\
 &
  Third Party Adveritising &
  Product Personalization &
  1 &
   \\
 &
  Product Personalization &
  Third Party Adveritising &
  4 &
   \\
 &
  Third Party Adveritising &
  Developers Advertising &
  3 &
   \\
 &
  Developers Advertising &
  Third Party Adveritising &
  2 &
   \\
 &
  Third Party Adveritising &
  Other Purposes &
  0 &
   \\
 &
  Other Purposes &
  Third Party Adveritising &
  1 &
   \\
 &
  Product Personalization &
  Developers Advertising &
  1 &
   \\
 &
  Developers Advertising &
  Product Personalization &
  1 &
   \\
 &
  Product Personalization &
  Other Purposes &
  1 &
   \\
 &
  Other Purposes &
  Product Personalization &
  1 &
   \\
 &
  Developers Advertising &
  Other Purposes &
  1 &
   \\
\multirow{-30}{*}{Browsing History} &
  Other Purposes &
  Developers Advertising &
  1 &
  \multirow{-30}{*}{104} \\
\hline
 &
  \cellcolor[HTML]{EA4335}App Functionality &
  \cellcolor[HTML]{EA4335}Analytics &
  \cellcolor[HTML]{EA4335}9 &
   \\
 &
  Analytics &
  App Functionality &
  3 &
   \\
 &
  App Functionality &
  Third Party Adveritising &
  4 &
   \\
 &
  Third Party Adveritising &
  App Functionality &
  0 &
   \\
 Sensitive Info &
  \cellcolor[HTML]{F4CCCC}App Functionality &
  \cellcolor[HTML]{F4CCCC}Product Personalization &
  \cellcolor[HTML]{F4CCCC}5 &
  56
   \\
 &
  Product Personalization &
  App Functionality &
  1 &
   \\
 &
  App Functionality &
  Developers Advertising &
  2 &
   \\
 &
  Developers Advertising &
  App Functionality &
  1 &
   \\
 &
  \cellcolor[HTML]{F4CCCC}App Functionality &
  \cellcolor[HTML]{F4CCCC}Other Purposes &
  \cellcolor[HTML]{F4CCCC}5 &
   \\
 &
  Other Purposes &
  App Functionality &
  2 &
   \\
 &
  Analytics &
  Third Party Adveritising &
  3 &
   \\
 &
  Third Party Adveritising &
  Analytics &
  0 &
   \\
 &
  Analytics &
  Product Personalization &
  2 &
   \\
 &
  Product Personalization &
  Analytics &
  4 &
   \\
 &
  Analytics &
  Developers Advertising &
  2 &
   \\
 &
  Developers Advertising &
  Analytics &
  1 &
   \\
 &
  Analytics &
  Other Purposes &
  2 &
   \\
 &
  Other Purposes &
  Analytics &
  1 &
   \\
 &
  Third Party Adveritising &
  Product Personalization &
  0 &
   \\
 &
  Product Personalization &
  Third Party Adveritising &
  3 &
   \\
 &
  Third Party Adveritising &
  Developers Advertising &
  0 &
   \\
 &
  Developers Advertising &
  Third Party Adveritising &
  1 &
   \\
 &
  Third Party Adveritising &
  Other Purposes &
  0 &
   \\
 &
  Other Purposes &
  Third Party Adveritising &
  1 &
   \\
 &
  Product Personalization &
  Developers Advertising &
  1 &
   \\
 &
  Developers Advertising &
  Product Personalization &
  0 &
   \\
 &
  Product Personalization &
  Other Purposes &
  2 &
   \\
 &
  Other Purposes &
  Product Personalization &
  1 &
   \\
 &
  Developers Advertising &
  Other Purposes &
  0 &
   \\
\multirow{-25}{*}{Sensitive Info} &
  Other Purposes &
  Developers Advertising &
  0 &
  \multirow{-25}{*}{56} \\ 
\bottomrule
\end{longtable}
\twocolumn
}

\newcommand{\privacyLabelExample}[0]{
\begin{figure*}[!t]
    \centering
    \fbox{\includegraphics[width=0.72\linewidth]{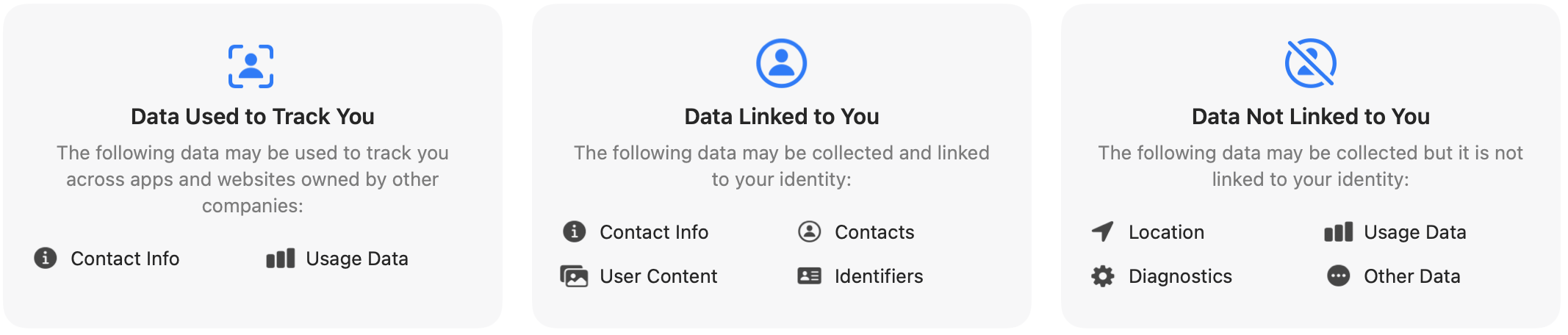}}
    \hspace{0.2in}
    \fbox{\includegraphics[height=1.25in]{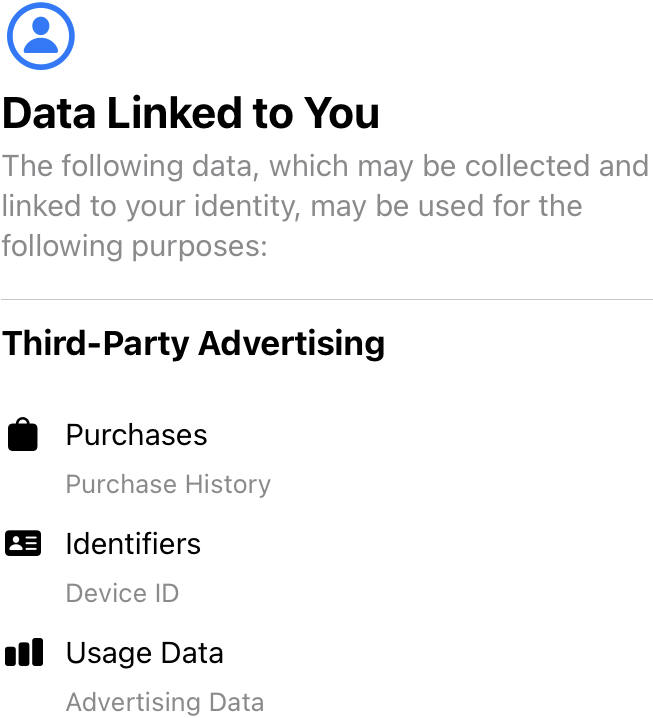}}
    \caption[Privacy Label With Three Privacy Types]{\label{fig:privacy-label-example}\label{fig:privacy-label-detail-example}
    {\em (left)} An illustrative example of a privacy label from the Apple App Store, and {\em (right)} an illustrative example of the privacy label details from the Apple App Store. The details display the \pus~ for the data collection and the detailed information about the \dts~ collected.}
\end{figure*}
}

\newcommand{\privacyLabelDetailsExample}[0]{
\begin{figure}[!t]
\centering
\fbox{\includegraphics[width=0.8\linewidth]{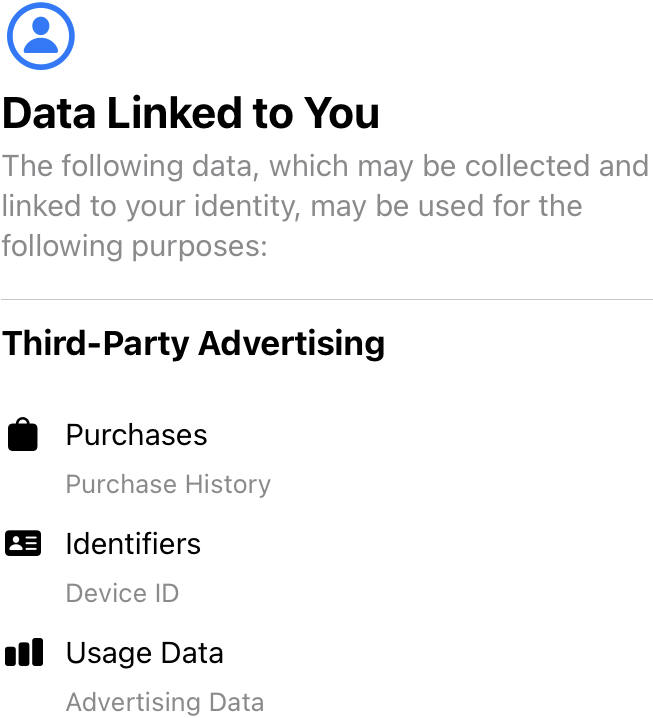}}
\caption[Privacy Label Details]{\label{fig:privacy-label-detail-example}
An illustrative example of the privacy label details from the Apple App Store. The details display the \pus~ for the data collection and the detailed information about the \dts~ collected.
}
\end{figure}
}

\newcommand{\anatormyOfLabel}[0]{
\begin{figure*}[!t]
\centering
\includegraphics[width=.8\linewidth]{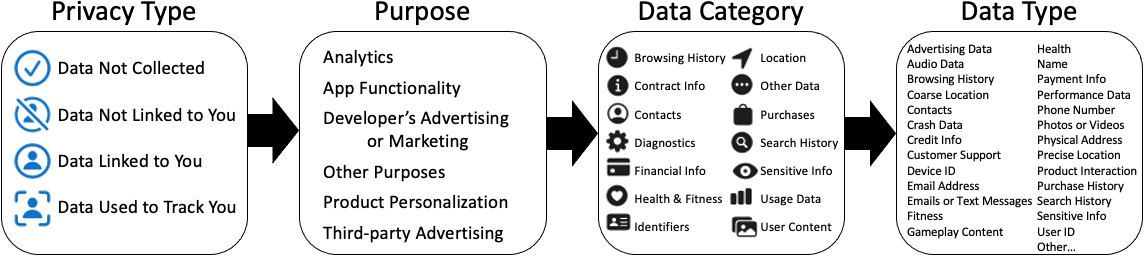}
\caption[Anatomy of a Privacy Label]{\label{fig:privacy-label-anatomy}
A privacy label consists of four hierarchical levels of information. The \emph{Privacy Type}  broadly identifies how the data collected will be used. The  \emph{Purpose} provides more detail on how each data type is used. The \emph{Data Category} is a categorization the \emph{Data Type} which is a detailed description of the type of data collected.
}
\end{figure*}
}

\newcommand{\privacyLabel}[0]{
\begin{figure*}[ht]
  \begin{subfigure}{0.49\textwidth}
    \centering
    \includegraphics[width=0.7\linewidth]{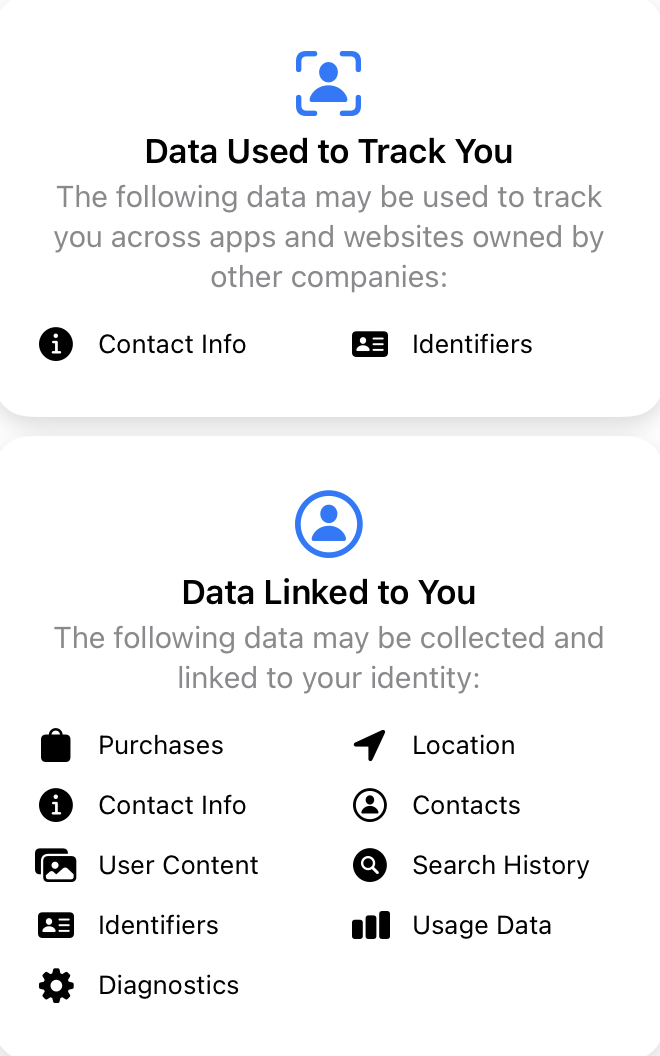} 
    \caption{Privacy Label.}
    \label{fig:privacy-label-image}
  \end{subfigure}
  \begin{subfigure}{0.49\textwidth}
    \centering
    \includegraphics[width=0.7\linewidth]{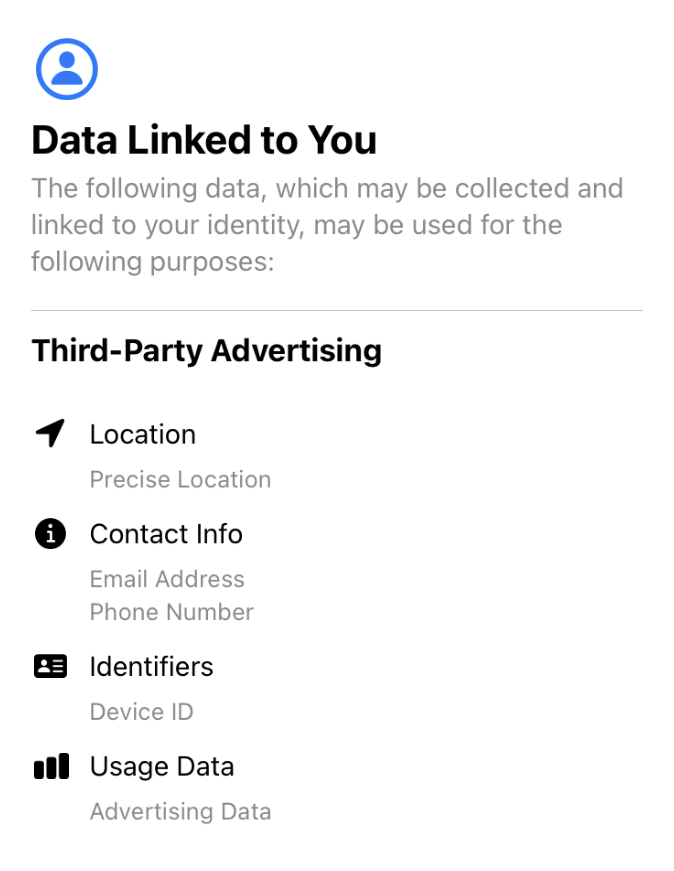}
    \caption{Privacy Label Details.}
    \label{fig:privacy-label-details}
  \end{subfigure}
  \caption{App Store Privacy Label.}
  \label{fig:privacy-label}
\end{figure*}
}

\newcommand{\privacyTypesVennDiagram}[0]{
\begin{figure}[!t]
\centering
\input{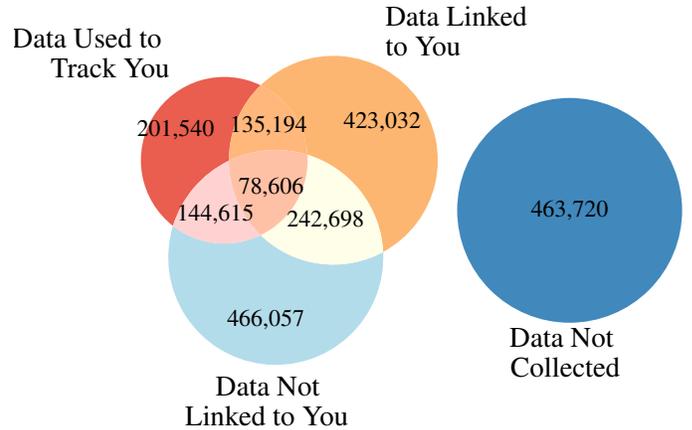}
\caption[Privacy Type Venn Diagram]{A Venn diagram of the number of apps in each of the four \emph{Privacy Types}. \emph{Data Not Collected} is mutually exclusive to the other three \emph{Privacy Types}.}\label{fig:privacy-types-venn}
\end{figure}
}

\newcommand{\dataTypeByPrivacyType}[0]{
\begin{figure}[!t]
\centering
\input{figures/data_type_by_privacy_type_ratio_no_right_spine.pgf}
\caption[Data Type By Privacy Type Ratio]{The ratios of the 32 \emph{Data Types} for each of three \emph{Privacy Types}. The denominator is the number of apps in the specific \emph{Privacy Type}.}\label{fig:data-type-by-privacy}
\end{figure}
}

\newcommand{\dataCategoryByPrivacyType}[0]{
\begin{figure}[!t]
\centering
\input{figures/data_categories_by_privacy_type.pgf}
\caption[Data Category By Privacy Type Ratio]{The ratios of the 14 \emph{Data Categories} for each of three \emph{Privacy Types}. The denominator is the number of apps in the specific \emph{Privacy Type}.}\label{fig:data-category-by-privacy}
\end{figure}
}

\newcommand{\purposesByPrivacyType}[0]{
\begin{figure}[!t]
\centering
\input{figures/purposes_by_privacy_type.pgf}
\caption[Purposes By Privacy Type Ratio]{The ratios of the six \emph{Purposes} for the \emph{Data Linked to You} and \emph{Data Not Linked to You} \emph{Privacy Types}. The denominator is the number of apps in the specific \emph{Privacy Type}.}\label{fig:purposes-by-privacy}
\end{figure}
}

\newcommand{\privacyTypeShifts}[0]{
\begin{figure}[!t]
\centering
\input{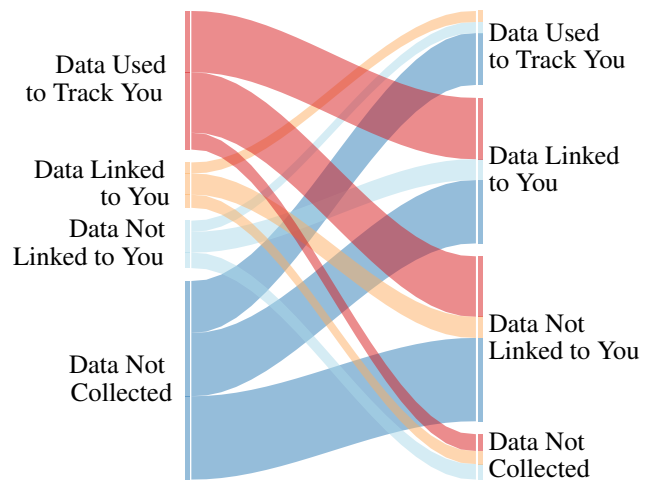}
\caption[Privacy Type Shifts]{Changes to the \emph{Privacy Types} associated with a \emph{posture} during our crawl. While the most significant shifts were \emph{from} \dnc, the highest shifts were observed \emph{to} \dly~and \dnly.}
\label{fig:privacy-type-shifts}
\end{figure}
}

\newcommand{\appSizeByPrivacyLabel}[0]{
\begin{figure*}[ht]
\centering
\input{figures/app_size_by_privacy_label_type.pgf}
\caption[App Size By Privacy Label Type Ratios]{The ratios of app sizes for each of the four \emph{Privacy Types}. The denominator is the number of apps with the designated app size that have a privacy label. Apps that are larger in size are more likely to collect data, including data used to track and linked to users. }
\label{fig:app-size-by-privacy}
\end{figure*}
}

\newcommand{\appCostsByPrivacyLabel}[0]{
\begin{figure*}[!t]
\centering
\input{figures/app_costs_by_privacy_label_type.pgf}
\caption[App Costs By Privacy Label Type Ratios]{The ratios of app costs for each of the four \emph{Privacy Types}. The denominator is the number of apps with the designated app cost that have a privacy label. Free apps are more likely than paid apps to collect data, including data used to track and linked to users.\label{fig:app-costs-by-privacy}}
\end{figure*}
}

\newcommand{\contentRatingByPrivacyLabel}[0]{
\begin{figure*}[!t]
\centering
\input{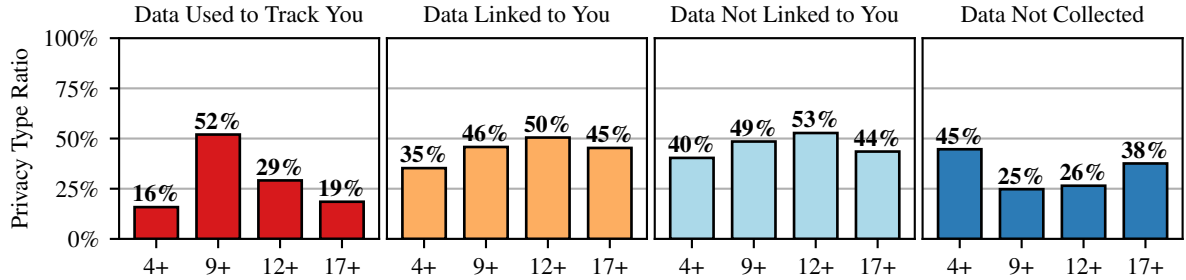}
\caption[Content Rating By Privacy Label Type Ratios]{The ratios of content ratings for each of the four \emph{Privacy Types}. The denominator is the number of apps with the designated content rating that have a privacy label.\label{fig:content-rating-by-privacy}}
\end{figure*}
}

\newcommand{\ratingsCountByPrivacyLabel}[0]{
\begin{figure*}[!t]
\centering
\input{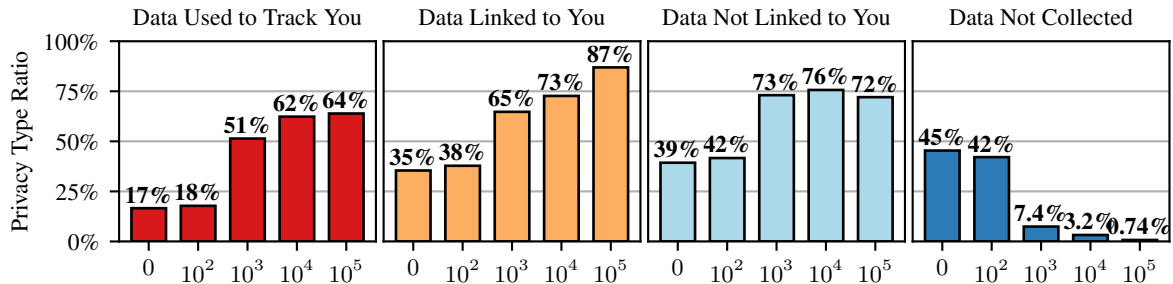}
\caption[Rating Counts By Privacy Label Type Ratios]{The ratios of the rating counts for each of the four \emph{Privacy Types}. The denominator is the number of apps with the designated rating counts that have a privacy label. Apps with a larger number of user ratings are more likely to collect data, including data used to track users. Ratings counts are not localized metadata and apps with low ratings counts in the US region may have higher counts elsewhere.\label{fig:ratings-counts-by-privacy}}
\end{figure*}
}

\newcommand{\releaseDateByPrivacyLabelMagnitude}[0]{
\begin{figure*}[ht]
\centering
\input{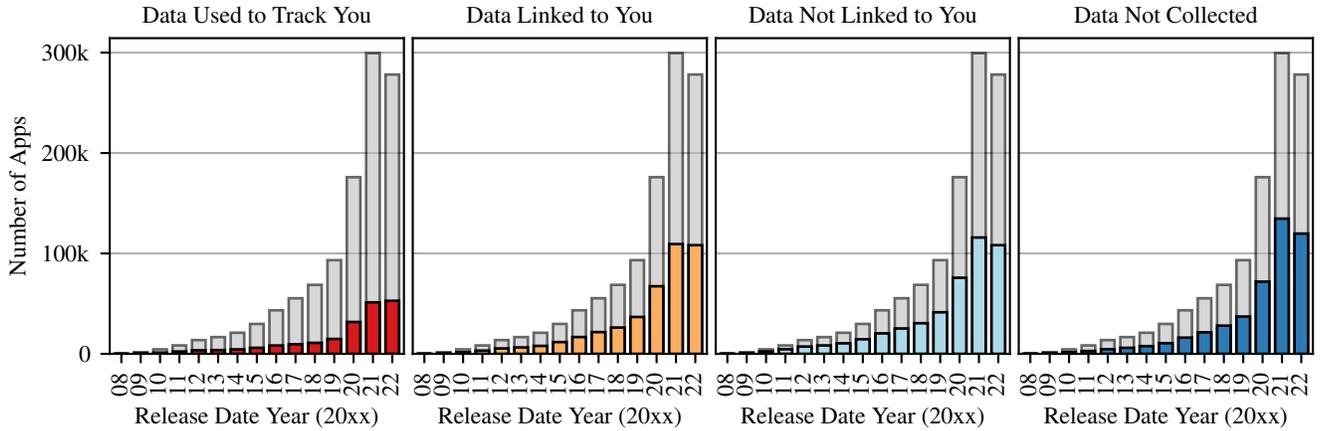}
\caption[Release Date By Privacy Label Type Magnitude]{The number of apps released during a given year for each of the four \pts. The gray bars show the total number of apps with privacy labels released in that year. The collection window includes apps through October 25, 2022.}\label{fig:release-date-by-privacy}
\end{figure*}
}

\newcommand{\appStoreGenreByPrivacyType}[0]{
\begin{figure*}[ht]
\centering
\input{figures/genre_by_privacy_label_type_ratio_no_right_spine.pgf}
\caption[App Genre By Privacy Label Type Ratios]{The ratios of top apps in app store genres for each of the four \emph{Privacy Types}. The denominator is the number of apps with the designated app store genre that have a privacy label. This includes only apps placed in the top in genre categories.}\label{fig:app-genre-by-privacy} 
\end{figure*}
}

\newcommand{\firstlabelswithupdate}[0]{
\begin{figure}[!t]
\centering
\input{figures/first_compliance_forced.pgf}
\caption[First Compliance]{The changes in \emph{Privacy Types} observed during the adoption of privacy labels by existing apps. App store policy enforced the addition of a privacy label to existing apps with their next version update. Apps that voluntarily added a label, without first releasing a version update, included more details about their data collection practices, while 50\% of the apps that added a label with a version update stated that they did not collect any data. \changes{Apps have multiple \emph{Privacy Types} per label, thus the percentages shown above will not sum to 100\%.}}
\label{fig:first-compliance}
\end{figure}
}

\newcommand{\categoryByPurposeCombinedHeatmap}[0]{
\begin{figure*}[!t]
\centering
\input{figures/category_purpose_combined_heatmap_no_title.pgf}
\caption[Category by Purpose Data Linked to You Heatmap]{The ratios of of \emph{Data Categories} by the reported \emph{Purpose} for the \emph{Data Linked to You} (orange) and \emph{Data Not Linked to You} (blue) \emph{Privacy Types}. }\label{fig:category-purpose-combined-heat} 
\end{figure*}
}

\newcommand{\categoryByPurposeLinkedHeatmap}[0]{
\begin{figure*}[!t]
\centering
\resizebox{\linewidth}{!}{\input{figures/category_purpose_linked_heatmap_no_title.pgf}}
\caption[Category by Purpose Data Linked to You Heatmap]{The ratios of of \emph{Data Categories} by the reported \emph{Purpose} for the \emph{Data Linked to You} \emph{Privacy Type}. }\label{fig:category-linked-heat} 
\end{figure*}
}

\newcommand{\categoryByPurposeNotLinkedHeatmap}[0]{
\begin{figure*}[!t]
\centering
\resizebox{\linewidth}{!}{\input{figures/category_purpose_not_linked_heatmap_no_title.pgf}}
\caption[Category by Purpose Data Not Linked to You Heatmap]{The ratios of of \emph{Data Categories} by the reported \emph{Purpose} for the \emph{Data Not Linked to You} \emph{Privacy Type}. }\label{fig:category-not-linked-heat} 
\end{figure*}
}

\newcommand{\privTypevsNumApps}[0]{
\begin{figure*}[!t]
\centering
\resizebox{\linewidth}{!}{\input{figures/privacy_type_per_run.pgf}}
\caption[Num. of apps vs. Num. Privacy Types]{A longitudinal view over the year-long collection period of the total number of apps and the total number of apps with privacy labels (compliant apps). For comparison, we also display the four \pts~ over the same period. Each data point represents a snapshot of the Apple App Store on that date.}\label{fig:priv-type-vs-num-apps} 
\end{figure*}
}

\newcommand{\privTypeChangeAcrossRuns}[0]{
\begin{figure}[!t]
\centering
\input{figures/privacy_type_change_acoss_runs.pgf}
\caption[Privacy Type Change Across Runs]{\emph{Posture} shifts to-and-from \emph{Data Not Collected}. Spikes observed in August and November appear to be due as a result of store-wide delays in updates ~\cite{app-submissions}.}\label{fig:priv-type-change-across-runs} 
\end{figure}
}

\newcommand{\UHaulScreenshots}[0]{
\begin{figure}[!t]
\centering
    \begin{subfigure}{0.8\columnwidth}
         \centering
         \includegraphics[width=\columnwidth]{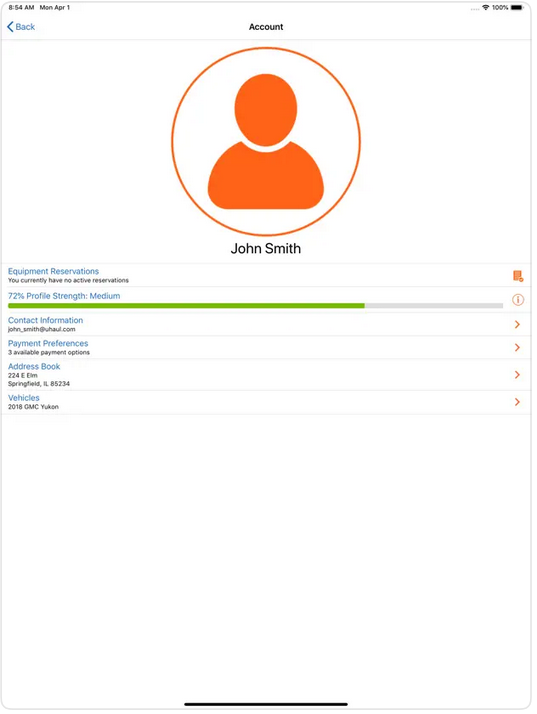}
         \caption{User details associated with an account on the app.}
         \label{fig:uhaul-account}
     \end{subfigure}
     \begin{subfigure}[]{0.8\columnwidth}
         \centering
         \includegraphics[width=\columnwidth]{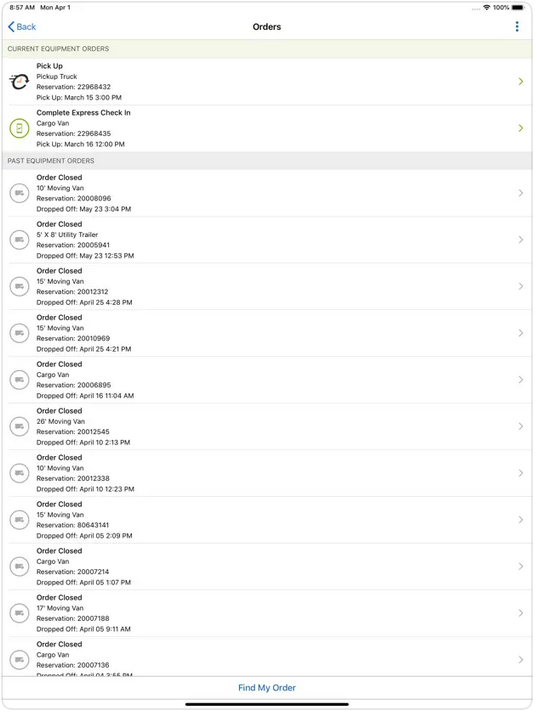}
         \caption{Purchase history of prior rentals made using an account.}
         \label{fig:uhaul-orders}
     \end{subfigure}
     \begin{subfigure}[]{0.8\columnwidth}
         \centering
         \includegraphics[width=\columnwidth]{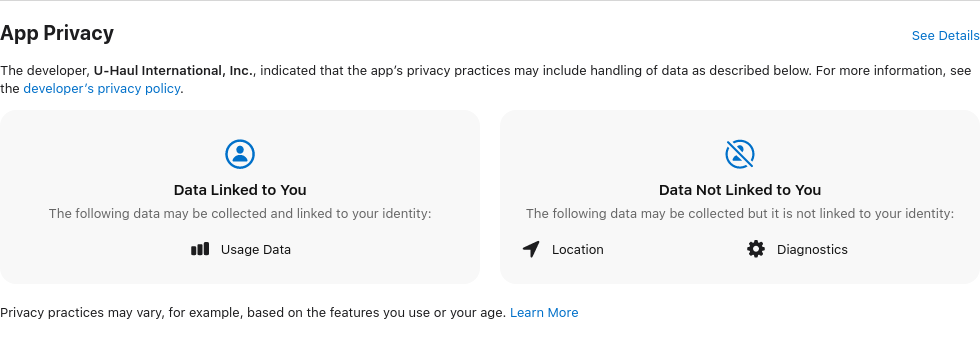}
         \caption{Updated U-Haul Privacy Label (since May 2022).}
         \label{fig:uhaul-orders}
     \end{subfigure}
\caption[U-Haul Screenshots]{\label{fig:uhaul-screenshots}
Screenshots of the U-Haul provided by the company for their app on the App Store~\cite{U-Haul:AppStore}.
}
\end{figure}
}

\maketitle
\thispagestyle{plain}
\pagestyle{plain}

\begin{abstract}

In December of 2020, Apple started to require app developers to self-report privacy label annotations on their apps indicating what data is collected and how it is used.
To understand the adoption and shifts in privacy labels in the App Store, we collected nearly weekly snapshots of over 1.6 million apps for over a year (July 15, 2021 -- October 25, 2022) to understand the dynamics of privacy label ecosystem.
Nearly two years after privacy labels launched, only 70.1\% of apps have privacy labels, but we observed an increase of 28\% during the measurement period. 
Privacy label adoption rates are mostly driven by new apps rather than older apps coming into compliance.
Of apps with labels, 18.1\% collect data used to track users, 38.1\% collect data that is linked to a user identity, and 42.0\% collect data that is not linked.
A surprisingly large share (41.8\%) of apps with labels indicate that they do not collect any data, and while we do not perform direct analysis of the apps to verify this claim, we observe that it is likely that many of these apps are choosing a \dnc~label due to being forced to select a label, rather than this being the true behavior of the app. 
Moreover, for apps that have assigned labels during the measurement period nearly all do not change their labels, and when they do, the new labels indicate more data collection than less. 
This suggests that privacy labels may  be a ``set once'' mechanism for developers that may not actually provide users with the clarity needed to make informed privacy decisions.

\end{abstract}
\section{Introduction}
\label{sec:intro}
Ubiquitous data collection embedded within modern web and mobile ecosystems has led to several interventions meant to empower users in managing their privacy. Notably, privacy policies offer free-text explanations of data collection and use but are often complicated and difficult for users to comprehend and apply to decision making~\cite{McDonald:2009,Cranor:2014}. %
Privacy nutrition labels (or {\em privacy labels})~\cite{kelley-nutrition-2009}  offer an alternative approach that is modeled after food nutrition labeling~\cite{nutrition-how-2022}, where a privacy nutrition label compactly describes the data collection and usage practices of a service via prescribed fields~\cite{emami-naeini-which-2021,kelley-nutrition-2009,kelley-privacy-2013}. %

In December of 2020, Apple mandated developers to self-report privacy labels for new and version-updated apps.
Apple's privacy labels (see \autoref{fig:privacy-label-example} and \autoref{fig:privacy-label-anatomy}) standardizes information such as the type of data collected by an app (e.g., \emph{Email Address}, \emph{Payment Info}, \emph{Precise Location}), the purpose of the collection (e.g., \emph{App Functionality}, \emph{Product Personalization}), and privacy-relevant aspects associated with the data collection (e.g., tracking users across apps and websites, linking to user identities, de-identifying or anonymization). In April of 2022, Google followed and announced its own form of privacy nutrition labels, {\em data safety labels}, for apps on the Play Store~\cite{noauthor-provide-nodate}.

\newcommand{\RQONE}[0]{What are the trends of privacy labels adoption?}
\newcommand{\RQTWO}[0]{What factors affect the adoption of privacy labels and how do they change?}
\newcommand{\RQTHREE}[0]{How are different dimensions of apps (based on app metadata) selecting privacy labels?}

We conducted more than a year-long (July 15,  \changes{2021} -- October 25, 2022) longitudinal analysis of over 1.6 million apps in the iOS App Stores adoption to understand how developers are selecting and changing privacy labels. Importantly, we focus only on the adoption and dynamics of the privacy labels ecosystem, rather than verifying their accuracy. We seek to answer the following research questions:
\begin{enumerate*}[%
label=\textbf{(RQ\arabic*)}, ref=\textbf{RQ\arabic*}]
\item \RQONE{}
\item \RQTWO{}
\item \RQTHREE{}

\end{enumerate*}

In answering RQ1, in October 2022, nearly two years after Apple's new policy, 70.1\% ($n=$ 1,110,448) of apps  have a privacy label, a 28\% increase since the start of the measurement period (670,547 vs. 1,110,448). However, this increase was primarily driven by new apps added to the store that are required to have a label rather than older, legacy apps  being updated or voluntarily adding labels.

When considering RQ2 and how labels change, it is again important to note that privacy labels are self-reported, entirely at the discretion of the developer, and not validated by Apple. As such, we observed different labeling depending whether the developers were forced to add labels, like a new or updated app, as compared to voluntarily adding labels. Fifty percent of developers forced to add labels when releasing a version update opted for \dnc~but only forty-one percent did so when voluntarily adding labels. Many developers may see privacy labels as an obstacle to the ultimate goal of adding an app to the store, rather than a mechanism to communicate about privacy practices. Moreover, we observed very few shifts in privacy labels (only 18,698 apps of more than 1M apps with labels). When labels do change, they tend to report more data collection.
\looseness=-1

Finally, we divided the apps based on a number of criteria in answering RQ3, including their (age-based) content rating, cost, popularity, and app-size. For example, when considering content-rating, 16\% and 52\% of apps with a content rating of 4+ and 9+, respectively, indicate that they collect data targeting users which can include children and would be subject to data collection and tracking standards in line with Children's Online Privacy Protection Act (COPPA). When comparing paid vs. free apps, more free apps use privacy labels that indicate data collection and tracking,
perhaps reflecting additional revenue streams from free apps in targeted advertising and/or selling user data. 

Our findings suggest that there are a number of factors in how developers select and update privacy labels in the Apple App Store. As labels are not validated, the true practices may vary significant from what is reported, especially for apps that explicitly report that they do not collect any data. Our measurement offers important baselines for future work in app analysis. Comprehensive and continual transparent measurements at scale provide important context accountability for developers. The trends and practices we present in our study offer important perspective into a particularly pressing issue regarding the adoption and use of declared data collection practices on the App Store, especially as privacy nutrition labels are becoming the {\em de facto} standards across the mobile app marketplaces beyond just Apple.

\section{Related Work}%

Researchers have gathered evidence of online services data collection behaviors via longitudinal measurements across various platforms~\cite{Lerner:2016, Englehardt:2016, Acar:2014, Pantelaios:2020, Kollnig:2021, Seneviratne:2015, Ren:2018, Khan:2018}. Analysis of mobile apps showed sensitive data, including PII~\cite{Ren:2018}, is collected and shared with third parties without user consent~\cite{Kollnig:2021, Bui:2022}. Data collection practices were prevalent across measurements of apps in different geographic regions~\cite{Shen:2021}, categories~\cite{Wang:2020,Feal:2020}, price brackets~\cite{Seneviratne:2015,Han:2019,Han:2020}, and app markets~\cite{Kollnig:2022,Wang:2018,Mehrnezhad:2020}.

Both Android and iOS require all applications to use install and/or runtime permissions~\cite{AndroidPermissions:2022, ApplePermissions:2022}, different from privacy labels. Prior research showed low attention and comprehension for install-time permissions~\cite{Felt:2012}. Felt et al.~\cite{Felt:2011} further found evidence  over-permissioning~\cite{Felt:2011}. Reardon et al.~\cite{Reardon:2019} showed apps circumventing the permissions by gaining access to data using covert and side channels. %

Kelly et al.~\cite{kelley-nutrition-2009, kelley-standardizing-2010} developed a privacy label to describe how websites collect, use, and share users' personal information, and later extended~\cite{kelley-privacy-2013} in the design of ``Privacy Facts.''
Subsequently, Emami-Naeini et al.~\cite{emami-naeini-ask-2020, emami-naeini-which-2021} developed and evaluated similar labels for Internet of Things (IoT) devices and found that users factor privacy risk perceptions into their purchase.  Over the years, multiple researchers have studied and provided recommendations on designing similar privacy notices from a variety of perspectives.~\cite{FTC:2011,Cranor:2012,Balebako:2015,kelley-nutrition-2009,kelley-standardizing-2010,kelley-privacy-2013,emami-naeini-ask-2020,emami-naeini-which-2021,Schaub:2015}.

In related work on Apples privacy labels, Li et al.~\cite{li-understanding-2022} interviewed 12 developers about selecting labels. Many developers misunderstood the process, leading to both  under-reporting and over-reporting data collection. Kollnig et al.~\cite{kollnig-goodbye-2022} evaluated 1,759 apps before and after they added a privacy label. They looked at instances of apps collecting an identifier for cross-device tracking, and inferred the impact that privacy labels had on such collection. They found apps adopting measures to circumvent Apple's detection of their tracking activity.
Zhang et al.~\cite{zhang2022usable} recently investigated the usability of Apple's Privacy Labels using semi-structured interviews with 24 iOS users. This study surfaced several potential concerns with the current implementation of privacy labels including clarity of the terse explanations provided by Apple for each label's meaning, and the lack of awareness that the labels are even included in the App Store listings. 

Garg et al.~\cite{Garg:2022} discovered that privacy label disclosures of sensitive information reduce app demand. %
Gardner et al. \cite{Gardner:2022} developed a  tool to assist developers to prompt  them of possible functionality that would require a privacy label.
Xiao et al.~\cite{Xiao:2022} analyzed 5,102 apps by checking the privacy labels against actual data flows,  discovering that 67\% of those apps failed to accurately disclose their data collection practices, particularly around the use of User ID, Device ID, and Location data.
Scoccia et al.~\cite{scoccia-empirical-2022} analyzed a small subset ($n=$ 17,312) of apps on the App Store. They captured \emph{two} snapshots of the subset of apps, seven months apart. They observed a \emph{decrease} in the number of apps that collect data for tracking purposes, but an \emph{increase} in overall data collection. Among our various additional results, we confirm their findings using 50 snapshots of 1.6M apps.%

Most relevant to our work is analysis by Li et al.~\cite{li-understanding-2022-1}. They collected weekly snapshots of privacy labels on the store between April 2 and November 5, 2021. They reported that only 2.7\% of apps during their collection period voluntarily added a privacy label, suggesting that inactive apps have little incentive to create privacy labels. They also observed 137,088 apps that created their first label in April 2021, and found that developers of these apps rarely updated privacy labels after creating one. We observed over 1M apps with labels that adopted a label at any point in our 66-week collection period before we could confidently corroborate their findings. Our analysis both verifies and significantly surpasses the work of Li et al. whose 31 weeks of analysis overlaps with less than half of our 66 week collection period. We offer a {\em detailed} analysis, and consider all criteria included in the privacy labels for each app, which they did not collect. We report not just on high-level privacy label data from an app's install page (see the boxes indicating \pts~on the {\em left} of \autoref{fig:privacy-label-anatomy}) but also include all privacy details about how data is used, i.e., \emph{Purpose}, and the granularity of the data collected, i.e. \emph{Data Types} (see {\em right} of \autoref{fig:privacy-label-anatomy}). We evaluate each label in a {\em complete} and {\em comprehensive} manner. We also report on a deeper set of app metadata to understand correlations with \emph{content rating}, \emph{user rating counts}, \emph{release dates}, \emph{app size}, and \emph{app price}.

\privacyLabelExample
\anatormyOfLabel

\section{Apple's Privacy Labels} 
In December 2020, Apple adopted app-based \emph{privacy labels}, making them mandatory for all new and updated apps on the App Store~\cite{inc-app-nodate}, and Google introduced {\em data safety labels} for the Play Store in April 2022~\cite{noauthor-provide-nodate}. Here, we focus exclusively on Apple's privacy label, which are similar in style to ``Privacy Facts'' by Kelly et al~\cite{kelley-privacy-2013}. 
The structure of Apple's privacy labels is hierarchical (see \autoref{fig:privacy-label-anatomy} for details) and are the combination of four sections of information. In the top level (to the left in the diagram) of the label hierarchy are four distinct \emph{Privacy Types}, three of which describe ways of using data. An app's privacy label may contain a combination of one, two, or all three of these \pts. The fourth \emph{Privacy Type}, entitled \emph{Data Not Collected}, is displayed with an image of a blue checkbox and indicates that the developer does not collect any data from this app. \emph{Data Not Collected} is mutually exclusive with the other three \emph{Privacy Types} (see  \autoref{fig:privacy-types-venn}). When an app adds a label with the \dnc~\pt, it states that it does \emph{not} collect \emph{any} data from the user, and therefore does not include other \emph{Privacy Types} or any \emph{Purposes}, \emph{Data Categories}, or \emph{Data Types}.

Among the three data collection \emph{Privacy Types}, the first, \emph{Data Used to Track You}, indicates that data collected may be used to track users across apps and websites owned by other companies, including sharing data with third-party advertising networks and data brokers. The second,  \emph{Data Linked to You}, indicates that data is collected and is linked to the user's identity. The third, \emph{Data Not Linked to You}, indicates that data is collected but is de-identified or anonymized and is therefore not linked to the user's identity. 

The next level down in the label hierarchy %
are the \emph{Purposes} which group data into six purposes for which data is collected, such as \emph{Third-Party Advertising}, \emph{Product Personalization}, or \emph{App Functionality}. \emph{Purposes}, which are only provided for the \emph{Data Linked to You} and \emph{Data Not Linked to You} \emph{Privacy Types}. The purpose for \emph{Data Used to Track You} \pt~indicates that the data is shared with third-parties for advertising or advertising measurement purposes, or to share data with a data broker, and hence does not require that an additional \emph{Purpose} be associated.

The \emph{Data Categories} are the next level, describing 14 categories of data collection, such as \emph{Contact Info}, \emph{Location}, and \emph{Purchases}. \changes{It is unlikely for an app to be  labeled so, but it is possible for all 14 \emph{Data Categories} to be listed under each of the three \emph{Privacy Types}.} 
And finally, at the bottom level (to the right in the diagram) are the \emph{Data Types},  which provide the most detailed grouping of the data collected into 32 descriptive types, such as \emph{Coarse Location}, \emph{Precise Location}, \emph{Gameplay Content}, or \emph{Emails or Text Messages}.

\autoref{fig:privacy-label-example} {\em (left)} provides an example of a privacy label as displayed. Each box reports the \pt, a short description of that type, and then within the box, each of the \dcs{}. When one clicks on a given category, a popup screen  is displayed, see \autoref{fig:privacy-label-detail-example} {\em (right)} with details about the \pus{} and specific \dts.

\privTypevsNumApps

\section{Data Collection Methods}
\label{sec:methods}

Beginning in July 15, 2022, each week we collect, store, and analyze data for an average of 1,598,134 ($SD=11,192$) apps. We enumerate all app URLs found on the Apple App Store (using the App Store's own site-map) and parse the privacy labels and metadata for each app. When parsing the data we extract: 
\begin{enumerate*}[label=(\roman*)]
\item app properties such as: 
    \begin{enumerate*}[label=(\alph*)]
        \item cost,
        \item size,
        \item developer,
        \item content rating,
        \item release date,
        \item and genre,
    \end{enumerate*}
\item what data is collected,
\item what is the purpose of the collection,
\item and how the data will be used.
\end{enumerate*}
The process proceeds in two stages: first, capturing the addition or removal of apps from the App Store; and second,  creating a new snapshot  for  retrieving the associated app metadata and privacy label information. 

First, a {\em cron} job will retrieve and parse the site-map from \url{https://apps.apple.com} weekly to extract  individual app URLs, which
are then insert the URLs into a database table for use in retrieving the privacy label information and associated app metadata. 
Second, using the full list of app URLs, a second process retrieves the app privacy labels and metadata. This is accomplished by parsing a JSON string embedded within each apps app-store page.
The privacy label are parsed directly from the HTML, which includes both the \emph{Privacy Types} and \emph{Data Categories}. 
To get the extended privacy label details, i.e., \emph{Purposes} and \emph{Data Types}, we perform a {\tt GET} request to the Apple catalog API at \url{https://amp-api.apps.apple.com/v1/catalog/}. %

The resulting dataset contains a total of 2,005,552 unique apps, with roughly 1.6M active per week. We had some data collection interruptions in late April through early September 2022 were our crawler was taking longer than a week to complete or was unable to get a full craw of the App Store. When this occurred we excluded this run. In total, we have 50 weeks of data during the 66 week data collection period.

\section{Limitations and Ethical Considerations}
\label{sec:limitations}

We sought to collect our data in an ethical manner by only accessing public  data on the Apple App Store web pages. We do not abuse any protocols or hidden APIs, but our measurement requests do incur an additional burden on Apple's servers. To alleviate that burden we place limits on the {\tt GET} requests. We request the web pages containing the app metadata in batches containing only 100 apps and initiate a sleep time of 10 seconds between batches. We additionally deploy an exponential backoff for errors to further reduce over requesting Apple's servers. At times, this led us to not being able to complete a crawl within a week. When this occurred, this data was excluded, and as such we are limited in what we can claim about weekly changes and instead report on changes between snapshots.

Privacy labels are \textit{self-reported} and selected by developers. They are neither regulated by Apple nor vetted by us. These values may \textit{not} accurately reflect actual data collection practices in the wild. %
However, we do provide examples where we believe labels are inaccurate (\autoref{sec:case-studies}) based on circumstantial evidence. We cannot report on inaccuracies at scale, though, given the nature of our measurement, but we do discuss possible methods to perform accuracy checks in \autoref{sec:discussion}. While we agree that an analysis of the accuracy of privacy labels is interesting, we argue that this is  beyond the scope of the analysis of this longitudinal measurement. 

Privacy labels were made mandatory on the App Store in December 2020. Since our collection of labels began in July 2021, we do not report on label adoption prior to that date. However, since less than half of the apps on the App Store had adopted privacy labels when we began our study, our year-long analysis importantly captured the adoption and changes of privacy labels from both, old and new apps.

The numbers presented below offer interesting insight into how apps posture their privacy and data collection practices to users, especially in the absence of any regulation from the App Store. They show us how apps wish to be perceived, and if they are incentivized to adopt and change these labels based on what users can see before choosing to install an app, regardless of the accuracy of the stated information.

\section{RQ1: Overall App Store Trends}\label{sec:results-app-store-trends}

\privacyTypesVennDiagram
\purposesByPrivacyType
\categoryByPurposeCombinedHeatmap
\dataCategoryByPrivacyType
\dataTypeByPrivacyType

In addressing RQ1, %
we consider each crawl of the app store and how many apps have labels and of what kinds (see \autoref{fig:priv-type-vs-num-apps}). There are an average 1,588,758 ($SD=$ 22,879) total apps in each crawl, and on average 10,989 ($SD=$ 11,063) newly published and 44,043 ($SD=$  20,031) removed apps in each crawl.
At the end of the data collection period, 70.1\% ($n=$ 1,110,448) of apps had a privacy label, which is a 28\% increase from the start where only 42.1\% ($n=$ 670,547) had labels. There still remains 39.5\% ($n=$ 621,908) of apps without privacy labels.

We find that 32.9\% ($n=$ 365,295) of apps have more than one \pt~(see \autoref{fig:privacy-types-venn}), and
7.1\% ($n=$ 78,606) of apps have all three \pts. The most common ($n=$ 242,698, 21.9\%) combination of \pts~ is the \dly{} together with \dnly, and \dnly~is the most common \pt{} overall with 42.0\% ($n=$ 466,057), followed closely by \dnc{} (41.8\%, $n=$ 463,720) of apps with labels. Note that \dnc{} is mutually exclusive with the other \pt, and an app reporting \dnc~ cannot report other privacy behavior. %

The reported prescribed \pus~ for the privacy labels are only provided for the \dly~ and \dnly~ \pts, as the purpose for the \duty~ is by definition only for targeted advertising. \autoref{fig:purposes-by-privacy} shows the ratios of the six \pus. The denominator for the ratios in this figure is the total number of apps labeled with the specific \pt. In the final snapshot of collection (October 25, 2022), the most common \pus~for collecting data linked to users' identities are \emph{App Functionality} ($n=$ 370,069; 87.5\%) and \emph{Analytics} ($n=$ 194,962; 46.1\%), and for collecting data not linked are \emph{App Functionality} ($n=$ 330,721; 71.0\%) and \emph{Analytics} ($n=$ 277,449; 43.1\%). These ratios of \pu~ were shown to be stable over the study period.

When we review the \dcs~ contained under each \pu, we find differences in the most common categories of data collection between \dly~ and \dnly. For example, under \dly~for the \pu~of \emph{App Functionality}, \emph{Contact Info} ($n=$ 292,137; 69.1\%) is the most common \dc, but under \dnly~ for the same \pu~ of \emph{App Functionality} it is \emph{Diagnostics} ($n=$ 192,007; 41.2\%). This suggests that while \emph{App Functionality} is often given as the reason for data collection, the category of data collected depends on whether that data is linked or not linked to a user's identity. A summary of the ratios of \dcs~by \pu~can be found in the heatmap of \autoref{fig:category-purpose-combined-heat}.

\autoref{fig:data-category-by-privacy} shows the ratios of the 14 \emph{Data Categories} for each of the three \emph{Privacy Types} as a heatmap. The most common \dcs~found under \duty~are \emph{Identifiers} ($n=$ 136,412; 67.7\%) and \emph{Usage Data} ($n=$ 117,246, 58.2\%). The \emph{Identifiers} category contains the \emph{User ID} and \emph{Device ID} \dts,  which are frequently used to track users for the placement of targeted advertisements. 
The most common \dcs~under \emph{Data Linked to You} are \emph{Contact Info} ($n=$ 307,403; 72.7\%) and again \emph{Identifiers} ($n=$ 274,063; 64.8\%). Very few apps report collecting and linked very sensitive user data such as \emph{Health \& Fitness} ($n=$ 18,683, 4.4\%) and \emph{Sensitive Info} ($n=$ 13,594, 3.2\%).

Apps have 2.2 ($SD=1.5$) \dcs~listed under \duty,  6.1 ($SD=6.2$) under \dly, and 3.6 ($SD=3.5$) under \dnly, on average. Viewed another way, there were a total of 4,692,465 instances of collected \dcs~on the app store, and 9.5\% ($n=$ 444,875) of them are \duty, 54.6\% ($n=$ 2,560,243) are \dly, and 36.0\% ($n=$ 1,687,347) are \dnly. A majority of data collected on the App Store is linked to users' identities and not anonymously collected.

Finally, at the most detailed level are the \dts. The specific detail about the type of data collected is important for broadly defined \dcs~ like \emph{Contact Info} and \emph{User Content}. For example, the \emph{Contact Info} category may include the \emph{Physical Address}, \emph{Email Address}, \emph{Name}, and \emph{Phone Number} \dts. For \duty~the most common \dts~collected are the \emph{Device ID} ($n=$ 116,500, 57.8\%) followed by \emph{Advertising Data} ($n=$ 86,223, 42.8\%) and \emph{Product Interactions} ($n=$ 73,265, 36.4\%).  For \dly~the most common are \emph{Email Address} ($n=$ 250,255, 59.2.4\%), \emph{Name} ($n=$ 255,148, 60.3\%), \emph{User ID} ($n=$ 220,790, 52.2\%), and \emph{Phone Number} ($n=$  209,272, 49.5\%). Refer to \autoref{fig:data-type-by-privacy} for full results.

\section{RQ2: Privacy Label Adoption and Changes} \label{sec:results-changes}
In this section we seek to answer RQ2, asking which factors affect the adoption of privacy labels and how and when does an app's privacy label change? We first consider the context in which privacy labels were adopted, either being required to because the app is new or has a version update, or if the developer simply updated metadata. Next we present details of if/when developers change previously posted labels. Finally, we provide some case studies to illuminate both adoption and changing behavior over the study period.

\subsection{Adoption Factors}
\firstlabelswithupdate
Recall that adding privacy labels is mandatory for new apps that are added to the store following December of 2020 and for any app that submits a version update. However, pre-existing apps can also update their privacy labels without updating the app (i.e., increasing the version number) by simply submitting a revision to their App Store page to include labels. 
\autoref{fig:first-compliance} presents the differences in how these two types of apps are labeled, those without a version update (voluntary addition of privacy labels) to those with a version update or new (forced adoption of privacy labels). There are markedly different distributions in privacy label types. A much larger share of apps that were forced to apply a privacy label choose \dnc{} as a label (50\% vs. 42\%). A similar trend is observed for other types. 
\looseness=-1

These results suggest that developers voluntarily opting into privacy labels may be taking a more genuine approach to selecting labels as they were not required to do so. In contrast, developers that were forced to add labels may have thought of the process as onerous and simply an obstacle to the end goal of adding or updating an app. They simply selected labels for the purpose of expediting the process. The divergence in distributions may imply that many of the labels are in fact speculative and applied conveniently for a large share of the apps in the store, particularly given that many of the apps that have privacy labels are new apps, as opposed to version updates or voluntary updates. Moreover, the first labels that are applied are critical; we do not find that many apps 18,698 made changes to the privacy label during our observation period (more details below). The initial privacy labels persist, and so the accuracy of the first labeling is key.

\privacyTypeShifts

\subsection{Privacy Label Shifts}
A total of 18,698 apps changed the \pts{} of their labels. As presented in\autoref{fig:privacy-type-shifts}, there are two common shifts. 
The largest is a privacy label shift from \dnc{} to any of the other three \pts{} ($n=$ 15,804). Most often this occurs from \dnc{} to \dnly{} ($n=$ 6,635), followed by \dly{} ($n=$ 5,040), and lastly, \duty{} ($n=$ 4,129). When an app developer decides to change a privacy label (which is rare) from \dnc, they are more likely to choose the less invasive of the \pts, e.g, \dnly. 
In contrast, the second common shift is moving from a more invasive label (\duty) to a less invasive one. \duty~($n=$ 11,027) shifted towards \dly~($n=$ 4,884) and \dnly~($n=$ 4,806), and a small number ($n=$ 1,377) moved from \duty{} to \dnc. Very few shifts ($n=$ 3,611) involved changes from a \pt{} where data is collected and/or tracked to \dnc.

We also measured how the data categories and their associated \pts{} changed for apps that had privacy labels. Recall that an app's privacy label is multi-leveled, beginning with the \pt{} (e.g., \dnc{} or \duty), and then under that, a developer can note the specific data categories being collected/tracked, as well as the purposes. 
Importantly, the \dcs{} and \pus{} can be changed without necessarily changing the \pts.

When we observe shifts in \dcs, developers almost always added new categories rather than removing existing ones. 
The most commonly added categories are \emph{Identifiers}, \emph{Usage Data}, and \emph{Diagnostics}, but these were also the most commonly removed \dcs. (See \autoref{tab:data-category-shifts-labeled}.)

\dcs{} can also be moved between \pts; for example, \emph{Contact Info} may have previously been of the type \dly{} and is now \duty. Similar to \dcs{} that were added, \dcs{} that were shifted most often between \pts{} were \emph{Usage Data}, \emph{Identifiers}, and \emph{Diagnostics}. Specific to these three \dcs, the most prominent shifts are towards \duty{} and \dly{} from \dnly. This is in contrast to other \dcs, where we more commonly observed a shift from \dly{} to \dnly{}. It is unclear if these shifts indicate that developers are making their \dcs{} more accurate or are attempting to obscure data collection practices of more sensitive data. That is, they are willing to note that data is collected, but do not want to indicate that it is actually linked to users.

We analyzed apps based on their App Store assigned \emph{genres} to determine if that impacted shifts in \dcs{} between \pts.
\emph{Games} engage in multiple \dcs{} shifts, more so than any other \emph{genres}. The \dcs{} that are most often shifted are \emph{Usage Data} ($n=$ 1043), \emph{Identifiers} ($n=$ 692), \emph{Diagnostics} ($n=$ 362), \emph{Location} ($n=$ 299), \emph{User Content} ($n=$ 112), \emph{Purchases} ($n=$ 112), and \emph{Other} ($n=$ 78) \dcs{} between \pts. 
The audience of mobile app game users is ``expansive''~\cite{DeloitteReport:2019, Schiff:Gaming:2020} and is a lucrative market for in-app advertising, which may explain why so many \dcs{} shifts occur~\cite{Dillon:ExchangeWire:2021}.

\dcs{} can also be assigned different \pus{} when labels change; for example, a label where \emph{Identifiers} were collected for \emph{App Functionality}  could change to\emph{Third Party Advertising}. Across different \dcs, labels most often change (or add) \pus{} to \emph{App Functionality}, \emph{Analytics}, and \emph{Product Personalization}. These shifts potentially indicate developers attempting to associate data collection practices with more beneficial purposes.

\dataCategoryShiftsLabeledTable

\subsection{Case Studies}
\label{sec:case-studies}

\paragraph{U-Haul}
U-Haul provides moving trucks, trailers, and self-storage rental services and lets users access their offering, reserve/pay for services, and manage their account through an iOS app~\cite{U-Haul:Website,U-Haul:AppStore}. %
U-Haul's App Store page~\cite{U-Haul:AppStore} includes screenshots of the app which show that a user account on the app includes their \textit{Contact Info} (Email Address, Physical Address), \textit{Financial Info} (Payment Info), and that the app saved previous rental orders (\textit{Purchases}) made using the account. While these screenshots are made available on the app's page by the developer themselves, their self-reported labels on the same page state otherwise. 

At the beginning of our measurement period, the U-Haul app had  a \dnc{} label, indicating that they did \textit{not} collect \textit{any} data from their users. This label is incorrect since the service relies on creating and maintaining user information, and offering purchases that are facilitated by gathering financial information from the user.

U-Haul retained this incorrect privacy label for a over a year during our measurement period, before they changed this composition in May 2022. The new label indicated that the app collects three data categories: (1) \textit{Usage Data} that is linked to users and used for \textit{Analytics}; (2) \textit{Diagnostics} data that is not linked to users (i.e., anonymized) and used for \textit{Analytics}; and (3) \textit{Precise Location} data that is not linked to users (i.e., anonymized) and used for \textit{Analytics} and \textit{App Functionality}. 
It is important to note that while the new label is more accurate, it is still incomplete. The screenshots and app description provided by the developer indicate that the app collects \textit{Contact Info}, \textit{Financial Info}, and \textit{Purchases}, none of which are declared in the privacy label~\cite{U-Haul:AppStore}. More alarming, the developer's privacy policy~\cite{U-Haul:PrivacyPolicy} states that they collect not just additional details about the user's \textit{Contact Info} (Phone Number), \textit{Identifiers} (Date of birth, Drivers License), and \textit{Sensitive Info} (Biometric information: ``Images of your face'').

\paragraph{First Command Bank}
Providing financial services aimed at American military families, the First Command Bank provides an app for their customers to access their checking, savings, and credit card accounts to check their balances, transfer funds, deposit checks, and pay bills~\cite{FirstCommand:AppStore,FirstCommand:Website}.
The screenshots and description of the app on the Ap Store indicate that the app requires users to create accounts where they can then link multiple bank and credit accounts, view and update them, and perform transactions. These interactions clearly require access to the user's personal \textit{Identifiers} along with their \textit{Financial Info} and \textit{Purchases} (i.e., transaction history).

At the beginning of our measurement period, the First Command Bank app had an elaborate privacy label that comprised 3 \pts{} and 7 \dcs{} -- (1) \textit{Contact Info} that is used to track users, (2) \textit{Purchases}, \textit{Financial Info}, \textit{Location}, \textit{Contact Info}, \textit{Identifiers}, \textit{Usage Data}, and \textit{Diagnostics}, that are collected in an identifiable manner, i.e., are linked to users, and (3) \textit{Financial Info}, \textit{User Content}, and \textit{Usage Data}, that are collected in an anonymized/unlinked manner~\cite{FirstCommand:Wayback}.

The app retained this label until October 2021 at which point they updated their label to a single, \dnc{} \pt{} indicating that they do \textit{not} collect \textit{any} data. The app and service, as indicated by their website, do not reflect any discernible privacy or functionality changes. This developers therefore knowingly changed their labels to reflect completely erroneous data collection practices.
At the end of our measurement period, the app retained this label, and was available for users to download. 

\contentRatingByPrivacyLabel
\ratingsCountByPrivacyLabel

\paragraph{Concours Avenir} 
A competitive application service based in France, Concours Avenir helps streamlines the application procedure for 7 major French engineering schools~\cite{ConcoursAvenir:Website}. The service lets students access education advice, testing services, information about schools. It helps them apply for and view results of the competition~\cite{ConcoursAvenir:AppStore}.
At the start of our measurement period, the app did not have a privacy label associated with it on the App Store. However, they adopted a label in October 2021. The first label was added without a corresponding version update, indicating that the developers voluntarily declared their data collection practices.

Their privacy label now indicates collection of \textit{Contact Info} to track users and link to users, \textit{Identifiers} linked to users, and finally \textit{Diagnostic} data not linked to users. The privacy label spans multiple \pts, in line with our findings (\autoref{fig:first-compliance}) that apps adopting labels without a version update where more likely to include more \pts.

\paragraph{Wordle!} 
Ranked \#2 in the \textit{Word Games} category on the App Store, the Wordle! app~\cite{Wordle:AppStore} provides an alternative to the popular game owned by New York Times~\cite{OriginialWordle:Website}. The app offers users an option to play the game multiple times a day and to compete against their friends.
At the beginning of our measurement period, the Wordle! app did not have a privacy label. The privacy policy provided by the app's developer, Lion Studios LLC, indicates that the service lets users create accounts and collects their registration and payment information~\cite{Wordle:PrivacyPolicy}.

The developer added a privacy label for the app in January 2022. Their privacy label adoption did not correspond with a version update, 
and their new label comprised two \pts, indicating that they collect data that is linked to users and used to track users. Their label mentions the collection of 7 \dcs, 5 of which are used to track users. This example also aligns with our previous findings that apps that voluntarily adopt a privacy label include more \pts{} and are more likely to indicate collection of data that is linked to users and used to track users.

\section{RQ3: App Metadata and Privacy Labels}\label{sec:results-metadata}

In this section, we consider app's metadata, such as content rating, app size, and price, as it relates to data collection practices in the privacy labels. Unless stated otherwise, all results are from the last snapshot (Run 50; October 25, 2022).

\paragraph{Content rating}
Apps' content rating describe  the age appropriateness of an app, e.g, is it for adults or children (i.e., 4+, 9+, 12+, and 17+), and are set by the developer based on Apple's guidelines~\cite{app-guidelines}. These content ratings should also comply with local policy requirements; for example, the Children's Online Privacy Protection Act (COPPA)~\cite{coppa-2013} in the US. 

Most apps with privacy labels have a content rating of ages 4+ ($n=$ 843,912; 75.9\%), while only 13.5\% ($n=$ 150,354) of apps with labels have a content rating of ages 17+. Of apps with privacy labels, 15.8\% ($n=$ 133,615) have a 4+ content-rating and are labeled with \duty~, and 35.2\% ($n=$ 297,536) are labeled with \dly. For the 17+ content rating, 18.5\% ($n=$ 27,834) were labeled with \duty~ and 45.3\% ($n=$ 68,121) with \dly. Refer to \autoref{fig:content-rating-by-privacy} for full details.

Apps with ratings of 4+ or 9+ that are {\em also} targeted at children would be subject to COPPA compliance in the US and require consent from parents to collect data that tracks minors under the age of 13. We did not perform a manual review of these apps to determine if they  are actually directed at children, but the fact that so many apps with a 4+ or 9+ content rating track data and would be available to children under the content rating guidelines (such as those used by parental controls) is problematic and worthy of further investigation. 
\appStoreGenreByPrivacyType
\releaseDateByPrivacyLabelMagnitude

\paragraph{Rating count} 
Users can rate apps on a five-star scale in the App Store (even without leaving a review), and the magnitude of ratings offers a reasonable proxy for app  popularity. Apps with over 100,000 ratings collect on average 19.25 ($SD=11.50$, $M=17$) \dcs~ per app, while apps with fewer than 100,000 ratings only collect on average 7.24 ($SD=6.86$, $M=5$) \dcs~ per app. Furthermore, 63.9\% ($n=$ 430) of apps with over 100,000 ratings were labeled with \duty~ and 86.9\% ($n=$ 585) with \dly. This is a noticeably higher proportion of data collection for tracking and linking purposes than apps with fewer ratings. For comparison, for apps with 100--1,000 ratings, only 17.8\% ($n=$ 195,622) are labeled with \duty~ and 37.8\% ($n=$ 415,698) with \dly. (See \autoref{fig:ratings-counts-by-privacy}.) These results indicate that more popular apps are more likely to report tracking and linking user data. Apps with more users may be monetizing their popularity~\cite{surveillance-capitalism}.%

\paragraph{Release date} 
Apps released in 2021 ($n=299,440$) were required to have a privacy label, and these apps make up 27.0\% of all apps with labels.
Older apps have made progress in coming into compliance, but at the end of the data collection a little over half (53.04\%; $n=$ 524,958) of apps released before the December 8, 2020 were privacy-label compliant. A significant share of older apps are still without labels, which may suggest that these older apps are no longer being supported with developer updates, or there is a lack of incentive for developers to make privacy label updates without being required to do so. Refer to \autoref{fig:release-date-by-privacy} for a detailed look at the number of apps with privacy labels by release year for each \pt.

\appSizeByPrivacyLabel
\appCostsByPrivacyLabel

\paragraph{App size} 
According to the privacy labels, larger apps collect and track more user data. This may be due to the fact that apps with larger footprints contain additional software libraries for this purpose. Additionally, game apps (as discussed below) are commonly tracking users and tend to require large downloads for graphics and other features. (See \autoref{fig:app-size-by-privacy}.)

\paragraph{App price} 
According to the self-reported privacy labels, free apps with in-app purchases are more likely to track and link to  users' identities, likely to generate revenues, then paid apps. Free apps report collect on average 7.30 ($SD=6.90$, $M=5$) \dcs~per app, while paid apps only collect on average 4.18 ($SD=4.53$, $M=3$) \dcs~ per app. This is in alignment with Scoccia et al.~\cite{scoccia-empirical-2022}, who made this comparison on a small subset of apps. These findings differ from the work of Han et al.~\cite{Han:2019, Han:2020} who investigated free and paid apps in the Android market based on inclusion of third-party advertising software, finding no differences between free and paid apps. Refer to \autoref{fig:app-costs-by-privacy} for full details.

\paragraph{Top chart app genre} 
The Apple App Store offers categorization of apps into genres, such as Social Networking, Games, etc., and 
we analyzed the \pt~ distributions for apps with labels in each of the genres. We found that Games apps are the most likely to track users, followed by Shopping, Music, Photo \& Video, and Social Networking. Shopping apps are most likely to link data to users' identities, followed by Magazines \& Newspapers, Finance, Lifestyle, Sports, and Social Networking apps. Shopping apps are the most likely to collect data not linked to users' identities followed by Magazines \& Newspapers, Entertainment, Weather, and Lifestyle. Refer to \autoref{fig:app-genre-by-privacy} for full details.

\paragraph{Location data} 
The collection and use of location data is a source of concern for users of mobile apps \cite{almuhimedi-your-2015}. We found that the \emph{Location} \dc~ was reported under \duty~ by 52,727 apps, under \dly~ by 127,464 apps, and under \dnly~ by 129,006 apps. In total, 254,145 apps self-report collecting location data. However they only constitute 22.9\% of the total number of apps with privacy labels. Kollnig et. al.~\cite{Kollnig:2021} found that 49.2\% of iOS apps request opt-in permissions for access to device location data. This suggests that either the apps that request access to location data are keeping the data local to the device, or that location data collection is under-reported on the App Store privacy labels.

\section{Discussion and Conclusion}
\label{sec:discussion}
This paper presents a large-scale, longitudinal evaluation of privacy labels in the Apple App Store. To accomplish this we collected and analyzed the privacy labels and other metadata for 1.6 million apps for over a year.  Through this analysis, we explored the following research questions, whose answers we summarize below:

\begin{itemize}
    \item \textbf{RQ1:} \textit{\RQONE} There is a steady increase in the number of apps with privacy labels, 70.1\% of apps have privacy labels as of October 2022, a 28\% increase since the start of the measurement period. While more than half of the apps report some form of data collection, there is still a large share of apps (41.8\%) indicate no data collection.  The most common purposes of data collection, either linked or not linked, is \textit{Analytics} and \textit{App Functionality}, where \textit{Identifiers} and \textit{Usage Data} are the most common data categories. 
    
    \item \textbf{RQ2:} \textit{\RQTWO} We identified a discrepancies in privacy labeling when app developers are forced, e.g., via a version update, to adopt a label compared to voluntarily applying a label, e.g., a meta-data update. App developers updating labels voluntarily are more willing to indicate broader classes of data collection then those forced, who are more likely to indicate data not collected. Moreover, when we observed 18,698 apps that changed their labels during the observation period, and when they do shift, most commonly they move towards more data collection. 
    
    \item \textbf{RQ3:} \textit{\RQTHREE} The metadata analysis indicated a number of inters-ting trends with respect to privacy labeling, including that more popular apps and larger apps self-report more data collection than less popular ones and smaller ones. There are also large numbers of apps with content-rating that applies to children performing data collection, which may be a violation of COPA if these apps specifically taget minors. We also show that free apps self-report collecting more data than paid apps. 
    
\end{itemize}
Based on these findings, we offer additional discussion and further interpretation below.

\paragraph{The high cost of free apps}
Many free apps are only free because they partake in the lucrative data collection and sharing practices. As we presented in Section~\ref{sec:results-metadata}, free apps collect, on average, three more categories of data than paid apps. When combined  with other app metadata properties, the contrast is more stark. For example, free apps with  more than 100,000 ratings (an indicator of audience size) collect on average 19.25 ($SD=11.50$, $M=17$) \dcs{} per app. The large audience increases the surveillance surplus, which  may make it harder for app sellers to resist collecting a wider range of data to increase profits. For instance, free \emph{Social Networking} apps with more than 100,000 ratings  collect an average of 23.64 ($SD=18.24$, $M=22$) \dcs{} per app.%

These findings differ somewhat from prior work by Han et al.~\cite{Han:2019, Han:2020}, where they used the inclusion of third-party libraries as a proxy for privacy behaviors, and compared free apps with their paid counterparts. While they found no clear difference in privacy behaviors between these pairs, our analysis looks at free and paid apps from an ecosystem-wide  perspective. At the very least, our findings confirm that the popular free \emph{Games}, \emph{Shopping}, and \emph{Social Networking} apps are also the top collectors of user data. 

\paragraph{Empowering users}
Ideally, iOS users would compare the privacy labels of apps with similar functionalities and select the app that best satisfies their personal privacy preferences. However, there is evidence  that users may not make such choices using install-time information.
In many ways, privacy labels function similarly to install-time permissions from Android~\cite{Felt:2012} as a pre-installation, one-time opportunity for review. And like install-time permissions, users are likely not sufficiently informed nor sufficiently motivated to take action at that time.  

A second challenge is that the privacy labels are integrated into the App Store, not the iOS device itself. There are no mechanisms for users to review already installed apps, other than go to the App Store and select each app one by one. And even if a user were to perform this action, there is no mechanism for them to become aware of changes or updates in the privacy labels for apps over time. It remains unclear how and where privacy labels are intended to assist users in making informed decisions about their apps.

\paragraph{A question of accuracy} 
There have been news reports~\cite{fowler-wapo-checked} and research~\cite{kollnig-goodbye-2022} about the inaccuracies found in App Store privacy labels. We illustrate in Section~\ref{sec:results-changes} that when apps are forced to add privacy labels due to a version change or as a new app they are more frequently providing the \dnc~\pt, potentially out of convenience, expediency, or deception. Furthermore, we describe in Section~\ref{sec:results-metadata} how 22.9\% of apps with privacy labels report collecting location data, but research~\cite{Kollnig:2021} has found that 49.2\% of apps request access to location data. These discoveries suggest under-reporting of data collection via privacy labels. When reviewing a few apps manually, we observed apps that have a \dnc~ label that conflicts with data collection practices outlined in their privacy policy.
Future research is required to determine the full scope and nature of the inaccuracies in the privacy label ecosystem.

Without additional oversight from Apple and without negative consequences for inaccurate labels, developers may simply never be motivated to create labels that accurately reflect their apps' data collection practices. A lack of credibility in the privacy label model will ultimately erode user confidence in the system and reduce the likelihood that market forces will force developers towers  more privacy preserving applications. 
Inaccurate labels are also harmful to users as  privacy expectations become misaligned with  apps' true data collection practices.

\paragraph{Communicating label updates to users}
When app developers update their privacy labels, as 18,698 apps did during our collection window, they tend to report \emph{more} data collection. Users who have already installed and are currently running these apps on their devices are not notified that the privacy label has been modified. We contend that a mechanism that provides users with an easily understandable description of the changes to the privacy labels should be provided to users upon label update. One suggestion would be to display a notification to users the next time they open the updated app. The notification would also be useful for those applications that were installed by users before they had privacy labels, but have since labels adopted.

\bibliographystyle{IEEEtran}
\bibliography{main.bib}

\end{document}